\documentclass[journal]{IEEEtran}
\usepackage{algorithm,algorithmic,amsbsy,amsmath,amssymb,epsfig,bbm,mathrsfs,fancyhdr,fancyvrb,subfigure,url,cite,multirow,xcolor}
\usepackage{array,bm}
\usepackage{graphicx}
\usepackage{subfigure}
\usepackage{epstopdf}
\usepackage{setspace}
\usepackage{colortbl}
\usepackage{cite}
\allowdisplaybreaks

\newcommand{\beq}{\begin{equation}}
\newcommand{\eeq}{\end{equation}}
\newcommand{\beqn}{\begin{eqnarray}}
\newcommand{\eeqn}{\end{eqnarray}}

\usepackage{amsmath}

\providecommand{\theoremname}{\textbf{Theorem}}
\providecommand{\propositionname}{\textbf{Proposition}}
\providecommand{\remarkname}{\textbf{Remark}}
\providecommand{\lemmaname}{\textbf{Lemma}}
\providecommand{\corollaryname}{\textbf{Corollary}}
\providecommand{\Definition}{\textbf{Definition}}

\newenvironment{definition}[1][Definition]{\begin{trivlist}
\item[\hskip \labelsep {\bfseries #1}]}{\end{trivlist}}

\begin{document}
\title{Incentive Mechanisms for Federated Learning: From Economic and Game Theoretic Perspective }
\author{Xuezhen~Tu,
       Kun~Zhu,~\IEEEmembership{Member,~IEEE,}
       Nguyen Cong Luong,
       Dusit Niyato,~\IEEEmembership{Fellow,~IEEE,}
       Yang~Zhang,~\IEEEmembership{Member,~IEEE,}
       and Juan Li,~\IEEEmembership{Member,~IEEE}
\thanks{X. Tu, K. Zhu, Y. Zhang and J. Li are with the College of Computer Science and Technology, Nanjing University of Aeronautics and Astronautics, Nanjing 210016, China (email: \{tuxz98, zhukun, yangzhang, juanli\}@nuaa.edu.cn).}
\thanks{N. C. Luong is with the Faculty of Computer Science, PHENIKAA University, Hanoi 12116, Vietnam (email: luong.nguyencong@phenikaa-uni.edu.vn).}
\thanks{D. Niyato is with School of Computer Science and Engineering, Nanyang Technological University, Singapore 639798 (email: dniyato@ntu.edu.sg).}
\thanks{K. Zhu is the corresponding author.}
}

\maketitle

\begin{abstract}
Federated learning (FL) becomes popular and has shown great potentials in training large-scale machine learning (ML) models without exposing the owners' raw data. In FL, the data owners can train ML models based on their local data and only send the model updates rather than raw data to the model owner for aggregation. To improve learning performance in terms of model accuracy and training completion time, it is essential to recruit sufficient participants. Meanwhile, the data owners are rational and may be unwilling to participate in the collaborative learning process due to the resource consumption. To address the issues, there have been various works recently proposed to motivate the data owners to contribute their resources. In this paper, we provide a comprehensive review for the economic and game theoretic approaches proposed in the literature to design various schemes for incentivizing data owners to participate in FL training process. In particular, we first present the fundamentals and background of FL, economic theories commonly used in incentive mechanism design. Then, we review applications of game theory and economic approaches applied for incentive mechanisms design of FL. Finally, we highlight some open issues and future research directions concerning incentive mechanism design of FL.
\end{abstract}


\section{Introduction}  \label{sec.intro}
\IEEEPARstart{I}{N} the past few years, we have witnessed the rapid development of machine learning (ML) in the field of artificial intelligence (AI) applications, such as computer vision, automatic speech recognition, natural language processing and recommendation system\cite{pouyanfar2018survey,Hatcher2018,Goodfellow-et-al-2016}. The success of these machine learning technologies, especially deep learning (DL), builds on large volume of data (i.e., big data). With the advent of the Internet of Things (IoT), massive data is collected by Internet connected smart devices with limited resources (e.g., smartphones, sensors, etc.). In most traditional ML technologies, the local data collected by smart devices need to be transmitted and processed at a cloud or data center to train effective inference models. However, this causes excessive computation and storage costs, and the smart devices also suffer from serious privacy leakage risk \cite{Zhu2018}.

To address the aforementioned issues, mobile edge computing (MEC) has been proposed as a solution where the computing and storage capabilities \cite{Mao2017} of end devices and edge servers are utilized to make model training closer to where data is generated. This paradigm has promoted the emergence of a new class of ML techniques that exploit the participation of numerous clients without exposing the original data. One most famous distributed machine learning framework is Federated Learning (FL). FL allows each client with data source, i.e., data owner, to train a local model, and a global model is obtained through model aggregation. This process will be repeated until an accuracy target of the model is reached. FL usually adopts client-server architecture for this interaction. By this way, it decouples ML from acquiring, storing, and training data in data centers, overcoming the limitations of traditional approaches.

In spite of the aforementioned great benefits of FL, it still faces several critical challenges. On the one hand, the data owners, i.e., clients, typically consume their resources, e.g., computing and communication resources, for the local training. This prevent self-interested clients from contributing their resources for FL model training, unless they can obtain sufficient economic compensation. On the other hand, some unreliable clients may perform undesirable behaviours, which affect the performance of global model of a FL task. In particular, the client may launch a poisoning attack \cite{Zhao2019,Zhang2021,Zhang2019} that sends malicious updates to mislead the global model parameters leading to the failure of current collaborative learning. In addition, the privacy can not be guaranteed, i.e, participants information is still in danger of exposure. For example, a generative adversarial network attack \cite{Chen2021} could be launched in FL, in which an adversary can pretend to be a participant engaging in the model training and learn other participants’ data.

As a result, designing an efficient incentive mechanism for FL is of crucial importance due to the following reasons. First, to train and update the local models, the clients need to consume the computing resource and network resource. Thus, the clients need to paid by reward or money to motivate them for their contributions. Second, the clients are different in behaviour, and thus they may be free to join and drop out the FL platform. This significantly impacts the accuracy and the delay of the training. Thus, a high incentive motivates more clients to join the FL platform and improves the training performance.

There are two challenging tasks for incentive mechanism designs in FL. The first is how to motivate and maintain more participants from the participants' perspective. That is, how to provide a participating opportunity that a high reward is allocated to the clients and the privacy of the clients can be guaranteed. The second challenge is how to evaluate the participants' contribution from the FL task publisher/platform's perspective. The demands of different FL task are various, thus how to maximize the sustainable operation of the federation while minimizing the incentive cost is challenging.

To address the aforementioned challenges, many work has been done from different aspects. Also, there are related surveys including economic and pricing models for edge computing \cite{qiu2021}, FL \cite{lim2020federated}, and incentive mechanism design for FL \cite{Zhan2021}, \cite{Zeng2021}, but they do not focus on how to adopt economic and game models to determine optimal interactions among buyers and sellers. To the best of our knowledge, there is no such a survey discussing the incentive mechanism design of FL specifically from economic and game aspects. This motivates us to deliver the survey with the aim of providing a  guide for the relevant researchers engaged in the study of FL incentive mechanism. We hope the reader will understand how economic and game theoretic approaches can be adopted to effectively design incentive mechanisms for FL.

The remainder of this paper is structured as follows. Section~\ref{sec.incentive.design.fl} reviews the background of FL such as the definition, architecture, advantages and incentive mechanisms in FL. Fundamentals of the economic and game theory for incentive mechanisms in FL are given in Section~\ref{sec.fundamentals}. In Section~\ref{sec.application.game.theory}, different game theoretic models used for incentive mechanisms in FL are discussed, including Stackelberg game, non-cooperative game, etc. Section~\ref{sec.application.auction.fl} reviews the auction approaches for incentive mechanisms in FL. Applications of contract and matching theory for incentive mechanisms in FL are discussed in Section~\ref{sec.application.contract.matching}. Section~\ref{sec.summary} summarizes potential future research directions. Finally, conclusions are drawn in Section~\ref{sec.conclusion}. The list of abbreviations frequently used in this paper is given in Table~\ref{tab:table1}.

\section{Incentive Mechanism Design for Federated Learning}     \label{sec.incentive.design.fl}    
    \subsection{Fundamentals of Federated Learning}     \label{subsec.fundamental.fl}
    The concept of FL was proposed by Google \cite{Yang2019}, \cite{Jakub2016,KonecnyMYRSB16,McMahan2016} which aims to build ML models based on distributed datasets across multiple devices while protecting data privacy. In general, FL refers to a collaborative learning method that enables participants to interact and cooperate with each other for generating a global model without data sharing. The FL process generally consists of three steps:
    \begin{itemize}
        \item \emph{Step 1 (Model initialization):} In this phase, the central server broadcasts the initialized global model  to the selected local clients that are known as participants.
        \item \emph{Step 2 (Local model update):} Each client trains its local model based on the shared global model and local dataset. After the training, the clients upload their local models to the server.
        \item \emph{Step 3 (Model aggregation):} Upon receiving the local models from the participants, the server uses model aggregation algorithms, e.g., Federated Averaging algorithm (FedAvg) which averages the parameters of the local models, to generate an updated global model. Then, the new global model is sent to participants for the next global training.

    \end{itemize}
    The step 2 and 3 are repeated until the global model reaches an accuracy target.

    \begin{table}[!t]
        \newcommand{\tabincell}[2]{\begin{tabular}{@{}#1@{}}#2\end{tabular}}
        \centering
        \small
        \caption{Major Abbreviations}
        \label{tab:table1}
        \begin{tabular}{|m{2cm}<{\raggedright}|m{4cm}<{\raggedright}|}
        \hline
        \rowcolor{black!30}\bf Abbreviation & \bf Description \\
        \hline
        BB & Budge Balance\\
        \hline
        CE & Computational Efficiency\\
        \hline
        DO(s) & Data Owner(s)\\
        \hline
        ED(s) & Edge Node(s)\\
        \hline
        EMD & Earth Mover’s Distance\\
        \hline
        IC & Incentive Compatibility\\
        \hline
        IR & Individual Rationality\\
        \hline
        KKT & Karush-Kuhn-Tucker\\
        \hline
        MDP(s) & Mobile Device Group(s)\\
        \hline
        MO(s) & Model Owner(s)\\
        \hline
        NE & Nash Equilibrium\\
        \hline
        SE & Stackelberg Equilibrium\\
        \hline
        SP(s) & Service Provider(s)\\
        \hline
        SV & Shapley Value\\
        \hline
        SWM & Social Welfare Maximization\\
        \hline
        WBB & Weak Budge Balance\\
        \hline
        \end{tabular}
    \end{table}

\begin{figure}[!t]
        \centering
        \includegraphics[width=2.9in,height=3.3in]{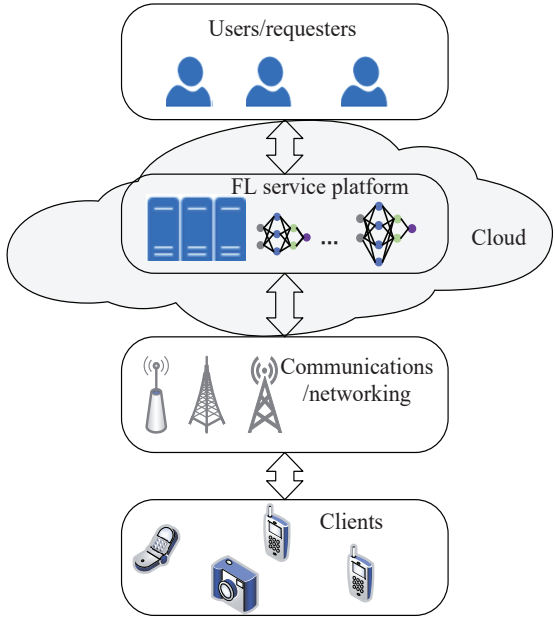}
        \caption{A general model of FL.}
        \label{fig:FL1}
        \end{figure}

    As an emerging distributed training framework, FL has been attracting great attentions. Multiple architectures of FL have been proposed based on different networks, such as client-server network and P2P network. However, for the works in this survey, the client-server network based architecture (as shown in Fig. \ref{fig:FL1}) is mainly considered, which consists of the following entities:
    \begin{itemize}
        \item \emph{Clients:} This layer is composed of devices, such as mobile devices, sensors, and vehicles. Their functions are to contribute resources, i.e, training data and computational resource to the FL process. Then, they act as the data owners (DOs) to train a model from the FL service platform, i.e., the model owner (MO), by using their collected data.
        \item \emph{Communications and networking:} This layer includes data communication and networking infrastructures for delivering the model parameters between clients and FL service platform.
        \item \emph{FL service platform:} This layer consists of sufficient data storage and computational resource to firstly aggregate the local model update from the clients, and then broadcasts the updated global model to all clients. Meanwhile, the platform provides FL services to service requesters or service users, e.g., through application interfaces.
        \item \emph{Service users/requesters:} This layer contains the users who request FL service.
    \end{itemize}

    With the above architecture, FL brings following benefits compared with conventional centralized learning scheme and individual learning scheme:
    \begin{itemize}
        \item \emph{Privacy protection:} Since the FL process does not involve the direct exchange or collection of raw data among participants, the privacy of participants is guaranteed to a certain extent. In a long-run, the privacy protection can better motivate participation, which further improves the performance of the FL system.
        \item \emph{Efficient model training:} For the large-scale FL-based training, only the model parameters (or model parameter updates) are sent to the model owner, which reduces the communication overhead. Besides, FL helps to solve the problem of training failure caused by the resources shortage.
        \item \emph{Lower local inference latency:} All participants can receive a better global model compared with self-training model. Besides, compared with the centralized model, in FL, making decisions can arise on local devices which leads to lower local inference latency.
    \end{itemize}

    \subsection{Incentive Mechanism Design for Federated Learning: Concepts, Definitions, and Motivations}
    \label{subsec.fl.concept.def.motiv}
    In the practical implementation of FL, participants may be reluctant to participate without receiving compensation due to the resource-consuming training model and the risk of privacy leakage. Additionally, there exists information asymmetry between the FL server and participants. Therefore, incentive mechanisms design is crucial for FL to encourage the clients for their participations and to reduce the potential adverse impacts of information asymmetry. In general, a desirable incentive can be characterized by the following  properties:
    \begin{itemize}
        \item \emph{Incentive compatibility (IC)/Truthfulness:} It means that each participant can obtain the optimal compensate when they truthfully report  their contributed resources and cost types. In other words, each of them can not improve their revenue by submitting the false information.
        \item \emph{Individual rationality (IR):} It means that the utility of each participant is non-negative when they join in FL.
        \item \emph{Budget balance (BB):} It means that the total payment for participants is no more than the given budget.
        \item \emph{Computation efficiency (CE):} It means that the incentive mechanism can complete the participant selection and reward allocation in the polynomial time.
        \item \emph{Fairness:} It means that the incentive mechanism achieves the property of fairness when some predefined fairness functions, e.g., contribution fairness \cite{Weng2018}, regret distribution fairness \cite{Yu2020},\cite{Yu2020a}, are minimized or maximized. Fairness is the key to keep FL sustainable. A fair incentive mechanism can optimally assign the reward to participants \cite{Zhang2020}.
    \end{itemize}
    \par To achieve the aforementioned properties, the incentive mechanism needs to be well designed. In the following, the definition of incentive mechanism and the main design phase are introduced in detail.
        \subsubsection{Definition of Incentive Mechanism in FL}
        An incentive mechanism in FL system can be simply formulated as a triplet $I=(\mathcal{P}, \mathcal{C}, \mathcal{R})$. $\mathcal{P}$ represents a set of potential FL participants which could provide contributed training resources, i.e., $\mathcal{Q}$, for the FL process. $\mathcal{C}$ denotes a method that is used to measure the contribution of each participant. $\mathcal{R}$ represents the reward assigned to each participant based on the measured contribution according to $\mathcal{C}$.

        Specifically, designing an incentive mechanism aims to determine the optimal participation level $\mathcal{Q}$ of participants and the optimal reward $\mathcal{R}$ to maintain the sustainability of FL. Optimization problems (e.g., utility maximization problem) are designed to obtain the optimal strategy.

        According to the above discussions, it can be known that the incentive mechanism includes two phases, i.e., contribution evaluation and reward allocation. Both of them are introduced in the following.

        \subsubsection{Contribution Evaluation}
        In FL, selft-interested DOs have a higher incentive to join FL if they receive higher rewards. However, this cause high incentive cost to the model owner. Thus, the contribution evaluation is important. Various methods have been proposed. In particular, the work in \cite{Huang2020} presented an exploratory analysis on honest DOs' contribution, malicious DOs' behaviors and the defense mechanism of attacks. The work in \cite{Liu2020b} adopted the attention mechanism to evaluate the contribution of gradients provided by DOs in vertical FL. This approach can obtain real-time contribution measurement for each DO with the high sensitivity on data quantity and data quality. Reference \cite{Nishio2020} presented an intuitive contribution evaluation method based on step-wise contribution calculation. In \cite{zhao2021efficient}, the authors proposed a reinforcement learning-based accurate contribution evaluation method. Particularly, the work in \cite{LvZL0THJL21} proposed a method named pairwise correlated agreement based on the idea of peer prediction to evaluate user contribution in FL without a test dataset, which perform it using the statistical correlation of the model parameters uploaded by users.

        However, the approaches in \cite{Huang2020,Liu2020b,Nishio2020} assumed that a trust central server will honestly measure the contribution of each DO, which lack transparency and may hinder the success of FL in practice. To address this issue, blockchain-based peer-to-peer payment system \cite{Ma2021,Liu2020} was proposed to enable SV-based profit allocation through consensus protocol which replaces the traditional third party. Also, to prevent malicious behaviors, the authors in \cite{Liu2020a} proposed a scoring rule based framework to incentivize DOs to upload their model updates in a trustworthy way.

        The major contribution evaluation strategies in existing FL system can be divided into the following categories:
        \begin{itemize}
          \item \emph{Self-report based Contribution Evaluation:} Self-report based contribution evaluation is the most straight-forward way, in which the DO actively reports the amount of its contributed resources to the MO. In the context of self-reporting, there are multiple metrics to evaluate DOs' contribution, such as the capacity of computational resource \cite{Sarikaya2020} and the data size \cite{Feng2019}.

          \item \emph{Shapley Value based Contribution Evaluation:} Shapley value \cite{Shapley2016} is a marginal method based on contribution which takes into account the impact of the participation order of DOs, to fairly evaluate their marginal contribution to the federation. This method is usually adopted in cooperative game. The SV of DO is denoted as:
              \begin{equation}
              \scriptsize
              \psi_{i}(N,v)=\sum_{\mathcal S\subseteq\mathbf{ \mathcal N}\backslash\{i\}}\frac{|\mathcal S|!(|N|-|S|-1) !}{N!}(v(\mathcal S \cup\{i\})-v(\mathcal S)),
              \end{equation}
         where $\psi_{i}(N,v)$ is the average marginal contribution of DO $i$  related to all possible subsets of federation, $\mathcal S$ represents the different cooperation modes in the large coalition $\mathcal N$, $v(\mathcal S)$ is the utility of model collaboratively trained by the subset $\mathcal S$. Recently, there has been many works on SV-based contribution evaluation of DOs \cite{Wang2019},\cite{Sim2020}, and its improvements \cite{Song2019}, \cite{Wang2020}.

          \item \emph{Influence and Reputation based Contribution Evaluation:} The influence of the client is defined formally as the effect of its contribution on the loss function of the FL model \cite{Richardson2020}. If the contribution is a model update (or data) the improvement of the loss function is achieved through applying the update (adding the data to the training set). The work in \cite{xue2021toward} defines a novel notion, Fed-Influence, to quantify the influence of each individual client in terms of model parameters, and it can well on both convex and non-convex loss functions. Reputation mechanism is introduced and combined with blockchain to select the reliable and trustworthy participants \cite{Roy2021,Kang2020,Feng2021,Lyu2020}. The reputation of the DO can be classified into direct reputation opinion and recommended reputation opinion, which can be  calculated using the subjective logic model \cite{Huang2018}.
        \end{itemize}

        \subsubsection{Reward Allocation}
        After evaluating the contribution of DOs, the MO should assign reward as the return of each DO to remain and enhance the participation level of high-quality DOs.
        \begin{itemize}
            \item \emph{Offered Reward:} It considers the case where the MO offers reward to the DO before training model, in which the reward can be determined according to the quality of the contributed resources \cite{Zhang2021a}, or through voting \cite{Toyoda2019}.
            \item \emph{Payoff Sharing:} It considers the scenario that the MO allocates the reward to the DOs when completing the FL task \cite{Bao2019}. Considering that payment delay may reduce the enthusiasm of participants in this scenario, the payoff sharing scheme \cite{Yu2020},\cite{Yu2020a} is proposed which can dynamically divide a given budget. The objective of this approach is to solve a value-minus regret drift optimization problem, which achieves contribution fairness, regret distribution fairness, and expectation fairness, simultaneously.
        \end{itemize}

\textbf{Summary:} In this section, we provide a brief introduction of FL that consists of a typical FL training process, general architecture, and its advantages. In addition, the fundamentals of incentive mechanism for FL are discussed, e.g., concepts, definitions and motivations. The next section presents some basic and fundamentals of economic and game models.

\section{Overview and Fundamentals of Economic and Game Theory for Incentive Mechanisms Design}
\label{sec.fundamentals}
\begin{figure*}[!t]
    \centering
    \includegraphics[width=7in,height=2.1in]{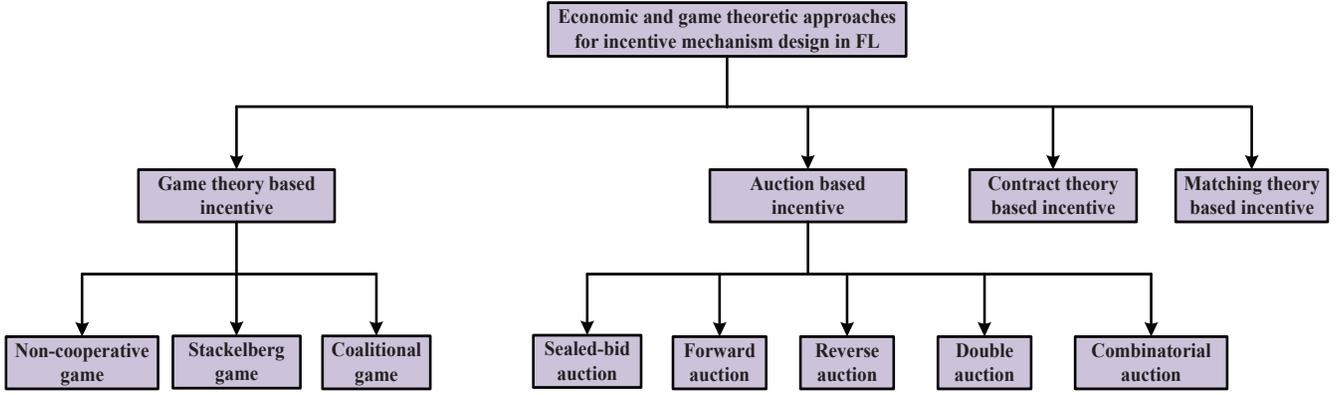}
    \caption{A taxonomy of economic and game approaches for incentive mechanism design in FL.}
    \label{fig:taxonomy}
    \end{figure*}

Economic and game theoretic approaches based incentive mechanism design for FL have received significant attentions. This section presents the fundamentals and background of the economic and game models commonly used for incentive mechanism design. A taxonomy of the economic and game models is provided in Fig.\ref{fig:taxonomy}.

    \subsection{Game Theory}
    Game theory is able to model the multi-participant interactive decision making problem in which the decision of a participant (i.e., player) potentially affects the decisions of other players. In the context of FL, the participants can be the MO (e.g., BS, MEC server, and SP) and the DO (e.g., mobile device, edge node, and users). In the following, we briefly present game theory-based incentive approaches which have widely been used to determine the best reward for participants of FL. To interpret the definition of the game, some terminologies are defined below.
        \begin{itemize}
            \item \emph{Player:} A player is a decision maker which chooses its actions, and is considered to be rational if it always plays in a way which maximizes its own payoff.
            \item \emph{Payoff:} The payoff refers to the benefits (e.g., profit, utility, interest) that a player can obtain from the game, which can be positive or negative.
            \item \emph{Strategy:} Player's strategy is a complete plan of actions that the player can choose to achieve a desired outcome. The payoff relies on not only the player's own actions but also the actions of others.
            \item \emph{Equilibrium:} Equilibrium means that each player in the game achieves the maximum utility they think, in which everyone has no an incentive to change its strategy. It reflects the stability of the game result.
        \end{itemize}

        \subsubsection{Non-cooperative Game}
        In a non-cooperative game, each player is considered to be selfish which only cares about the maximization of its own payoff rather than the social welfare of the FL system. In such games, there is not cooperation or agreements among players.

        Consider the pricing in a FL service market as an example. The mobile devices as sellers (i.e., DOs) can set the price for providing computational resources to the MO as buyer. A non-cooperative game among the competitive sellers can be modeled and defined as a triplet $ G=\left( P_i, u_i, i \in \mathcal{N}  \right)$, in which $\mathcal{N} $ represents $N$ competitive sellers, $P_i$ is a set of pricing strategies of player $i$, and $u_i$ is the payoff of player $i$ given $i$'s chosen strategy and the others' strategies. Let $p_i \in P_i$ be any possible strategy of player $i$.

        \begin{definition}{1:}
            \emph{ Denote $p_{i}^{*}$ as the best response of player $i$ to the chosen strategies of other $N-1$ players, i.e., $p_{-i}^{*}$. Then, the set of strategies $(p_{i}^{*},......,p_{N}^{*})$ is termed a Nash equilibrium (NE) of this game, meaning that no player can gain higher payoff by deviating its current strategy when the strategies of the other players remain the same, that is \cite{Nash1951}, }
            \begin{equation}    \label{eq.nasheq}
            u_i\left(p_{i}^{*}\,\, , p_{-i}^{*} \right) \geqslant u_i\left(p_i\,\, , p_{-i}^{*} \right) , \forall p_i\in P_i    .
            \end{equation}
        \end{definition}

        The inequality~(\ref{eq.nasheq}) implies that NE is a stable outcome of the game in which the sellers at the equilibrium have no an incentive to change their strategies since they cannot improve their payoffs unilaterally. Note that, there could be no NE or multiple NE in a game. This makes players difficult to predict the outcome of the game. Therefore, it is important to check the existence and uniqueness of the NE when using the non-cooperative game. It has been proven that a unique NE exists if the strategy space of each player is a convex set which is non-empty, closed, and bounded, and if its payoff function is a continuous and quasi-concave function \cite{Debreu1952,Fan1952,Glicksberg1952}. The existence and uniqueness of NE can also be proved by adopting the Kakutani's fixed point theorem \cite{Glicksberg1952} or the Brouwer's fixed point theorem \cite{Kakutani1941}. Because the non-cooperative game formulates the conflict relationship among self-interest players, the incentive problem among competitive participants of collaborative learning under limited budget was modeled as non-cooperative game \cite{Pejo2019}.

        The non-cooperative game assumes that information such as feature, payoff function and strategy of the players are the common knowledge among players, which is termed a complete information game. However, in practice, a player may not be fully aware of this information of other players, only knowing the occurrence probability of each type. Such a game is known as an incomplete information game. In FL market, the prior knowledge about Dos' reliability or reputation that helps allocate rewards may be unknown for the MO. A typical example is the Bayesian game \cite{Fudenberg1991} where the outcome of the game can be predicted using Bayesian game analysis. The equilibrium solution of such a game is Bayesian Nash equilibrium (BNE). Similar to NE in a complete information game, BNE can be obtained where each player selects a strategy that maximizes its expected payoff given its beliefs about the type and strategies of other players. Because the non-cooperative game models the conflict of selfish players, the game in the FL market where the DOs are competitive and the budget of the MO is limited can be formulated as the non-cooperative game. The corresponding Nash equilibrium allows DOs to have an optimal participation strategy. In the context of FL, it is used for the computation resource trading with competitive prediction service provider as proposed in \cite{Weng2020}.

        \subsubsection{Stackelberg Game}
        Stackelberg game \cite{Amir1999} is a sequential-move game in which the players acting as the leaders move first and then other players acting as followers move after observing leaders' moves. Therefore, it is also termed as leader-follower game \cite{Fudenberg1991}. The Stackelberg game aims to model multi-agent decision making processes and maximize the utility of both the leader and the followers given the leader's strategy.

        Consider again the scenario in Section \uppercase\expandafter{\romannumeral3-A} with two mobile devices as players, i.e., computational resource sellers. $P_1$ and $P_2$ denote the sets of the pricing strategies of players 1 and 2, respectively. Both mobile devices 1 and 2 aim to maximize its own utility $u_i\left(p_{1}\,\, , p_{2} \right), i\in 1,2$, where $p_1$ and $p_2$ are their chosen strategies from $P_1$ and $P_2$, respectively. Assume that player 1 chooses its strategy at stage 1, and thus acts as the leader. Player 2 chooses its strategy at stage 2, and acts as the follower. The optimization problem of the leader and the follower together form the Stackelberg game, and their solutions construct the Stackelberg equilibrium (SE).

        \begin{definition}{2:}
            \emph{Let $p_{1}^{*}$ and $p_{2}^{*}$ denote the solutions of the optimization problems of the leader and the follower, respectively. Then, the point $(p_{1}^{*} , p_{2}^{*})$ is the SE for the Stackelberg game if any $(p_1 , p_2)$ with $p_1 \geqslant 0$ and $p_2 \geqslant 0$, we have $u_1\left(p_{1}^{*}\,\, , p_{2}^{*} \right) \geqslant u_1\left(p_{1}\,\, , p_{2}^{*} \right) $ and $u_2\left(p_{1}^{*} , p_{2}^{*} \right) \geqslant u_2\left(p_{1}^{*} , p_{2} \right) $.}
        \end{definition}

        The backward induction method \cite{Aumann1995} is commonly used to derive the SE. In the above example, given $p_1$, the optimization problem of the follower is solved first to find $p_{2}^{*}$, and then the $p_{1}^{*}$ is obtained through substituting $p_{2}^{*}$ in the leader problem. Since player 1 takes advantage of knowing player 2 knows $p_1$, player 1 imposes a solution which will facilitate itself. Therefore, the utility of the leader at the SE is higher than that of the follower, called the first mover advantage. Accordingly, when reaching SE, the leader could achieve a payoff at least as high as the one obtained from the corresponding NE. This feature makes the Stackelberg game suitable for the incentive mechanism design in FL. For example, it allows the mobile devices (i.e., the followers) to determine optimal computational resource prices after knowing the amount of CPU resources that the MO (i.e., the leader) needs as proposed in \cite{Sarikaya2020} or the reward offered by the MO as proposed in \cite{Zhan2020a}.

        \subsubsection{Coalitional Game}
        In a cooperative game, players cooperate with each other with the aim of maximizing a common objective of the coalition. Moreover, enforceable contracts are made among the players. In this case, the players can coordinate strategies and reach an agreement on how to assign the total payoff to the players in a coalition. The objective of a coalitional game is to find a stable solution which ensures that the outcome of the game is immune to changes of groups of players (i.e., each player has no incentive to move from its current coalition to another coalition).

    \subsection{Auction}
    An auction is an economic mechanism, the goals of which are to allocation commodities (e.g., training data, computational resource, and bandwidth) and establish corresponding prices via a process known as bidding \cite{McAfee1987}. An auction consists of an explicit set of rules which determine resource allocation and prices of the basis from market participants \cite{McAfee1987}. There are following terminologies used in the auction:
    \begin{itemize}
        \item \emph{Bidder:} A bidder is a buyer who wants to purchase items in the auction. In the FL market, bidders can be model owners or FL service requesters.
        \item \emph{Seller:} A seller offers services or resources to the buyers. In FL markets, the sellers are typically data owners or clients who use their local data to train the shared model required by the model owner.
        \item \emph{Auctioneer:} An auctioneer is an intermediate agent which conducts the auction, i.e., price and winner determination. In many cases, the seller itself is the auctioneer.
        \item \emph{Price:} A price in an auction may be an asking price or a bidding price. The asking price is the price of a commodity that the seller wants to obtain, and the bidding price is the price that the buyer wants to pay for his requested commodity. Hammer price is the price at which the buyer and the seller agree to make a deal, i.e., final payment.
        \item \emph{Commodity:} An auction commodity refers to the object traded between a buyer and a seller. Each commodity has a value at which the
        buyer/seller wants to buy/sell. In FL markets, the commodity can be a data unit (a training data sample) or a computing resource unit that the data owner offers.
        \item \emph{Valuation:} In an auction, the valuation refers to monetary valuation of commodities. Different buyers and sellers may value commodities with different valuations depending on their preferences. The valuation of a participant can be private that is unknown to the other's participants or public that is known to the others.
        \item \emph{Utility:} The buyer's utility is the difference between its valuation of the requested commodity and its final payment. The seller's utility, i.e., revenue, refers to the total payment received from the buyers. In FL markets, the utility of the buyer, e.g., model owner, can be proportional to the accuracy of the global model and inversely proportional to the total payment that the model owner pays the data owners.
        \item \emph{Social welfare:} It refers to the sum of utilities of the users (i.e., both buyers and sellers) in an auction.
    \end{itemize}

    Auction mechanism has been widely applied in many fields, such as resource allocation in wireless systems \cite{Zhang2013}, secure data offloading \cite{Rayan2017}, and network security \cite{Luong2017}. For the rest of this section, we introduce the detail of auction types commonly applied to incentive mechanism design in FL.

        \subsubsection{Sealed-bid Auction}
        Different from the open-cry auction (e.g., English auction and Dutch auction), the bids of the buyers are open to each other during the auction, in a sealed-bid auction, the buyers submit sealed bids simultaneously to the auctioneer. Accordingly, no bidder can know the bidding information of others and cannot change its own bid. There three types of sealed-bid auctions.

        \begin{itemize}
            \item \emph{First-price sealed-bid (FPSB) auction:} The bidder with the highest bid is the winner who can receive the item and pays the highest bid.
            \item \emph{Second-price sealed-bid (SPSB) auction or Vickrey auction:} In this auction, the winner only pays the second-highest bid rather than the highest bid that it submitted \cite{LuckingReiley2000}. Since the winner pays the price less than its expected price, the Vickrey auction motivates buyers to bid truthfully. The auction thus achieves truthfulness.
            This feature enables Vickrey auction to be widely used for the incentive mechanism design in FL to prevent the misbehaviors of unreliable clients.
            \item \emph{Vickrey-Clarke-Groves (VCG) auction:} A VCG auction is a generalized Vickrey auction with multiple commodities. In the VCG auction, the commodities are allocated socially optimal manner, and the winner pays for the loss of the social value owing to winning the commodities. Such payment rule enables bidders to give their true value for the commodities. The VCG is thus a strategy-proof or truthful mechanism. In FL, VCG mechanism can be used to motivate IoT devices (the Dos) to report their true valuations to the network operator for maximizing social welfare as proposed in \cite{Kim2020}.
        \end{itemize}

        \subsubsection{Forward, Reverse and Double Auction}
        The auction mentioned above are classified as the forward auctions from the seller's side. Considering the buyer's side, there are reverse and double auctions. In particular, there are following definitions:
        \begin{itemize}
            \item \emph{Forward auction:} In the forward auction, multiple buyers submit their bids, i.e., bidding price, to compete for the requested items offered by one seller.
            \item \emph{Reverse auction:} In the reverse auction, multiple sellers submit their asks, i.e, asking price, to compete for selling the items to the single buyer. The reverse auction is often used together with other auction mechanisms, e.g., sealed-bid reverse auctions.
            \item \emph{Double auction:} In FL market, there may exists multiple MOs and multiple DOs, the double auction can be used to match the supply and demand. In a double auction, buyers and sellers simultaneously submit their bids, i.e., bidding price, and asks, i.e, asking price, to an auctioneer\cite{Daniel1993}. The auctioneer determines a price $p$, i.e., the transaction price, to clear the market, at which the asking prices from sellers are less than $p$ while the bidding prices from buyers are more than $p$. The transaction price is typically set as $p=\left(p_b + p_s \right)/2$, where $p_b$ is the bidding price of one buyer and $p_s$ is the asking price of one seller. The buyers receive resources, and the sellers gain the transaction price. The process is repeated until no more transactions occur or a predetermined end time achieves.
        \end{itemize}

        \subsubsection{Combinatorial Auction}
        In combinatorial auction, each bid of a buyer indicates a bundle of multiple commodities rather than an individual commodity \cite{Cramton2010}. Based on the information included in the bid as well as the capacity of commodities from sellers, the auctioneer determines the optimal allocation strategy as well as the winner of the auction. However, solving winner determination problem is a challenge for the combinatorial auction since the problem is generally NP-hard, which means that there is no polynomial-time algorithm to find the optimal allocation. There are many algorithms that have been proposed to obtain the approximate solutions for the problem, such as the Lagrangian relaxation approach \cite{Hsieh2010}. In FL, the combinatorial auction is used to allocate network operator's bandwidth to multiple FL SP as proposed in \cite{Xu2021}.

    \subsection{ Contract and Matching Theory}
    Contract theory \cite{Bolton2005} and matching theory \cite{Roth1992} have been regarded as two powerful tools to model the dynamic and mutually beneficial relations among different types of rational and selfish agents. In particular, they can effectively deal with the high dynamics of trading market, selfish and competitive participants.
    In the following, the brief introductions of contract theory and matching theory which have been used to design incentive mechanism in FL is presented.

        \subsubsection{Contract Theory}
        Contract theory is an economical theory that regards all transaction and institutions as a kind of contract \cite{Huang2020}. It is widely used where the asymmetric information is available between employer and employees (i.e., the futures of an employee is not known exactly to the employer). In the FL market, since the employees are selfish and they may not reveal their true bids as well as the property of privacy protection  of FL mechanism, there exists information asymmetry. Contract theory can design the optimal contract to reduce the moral hazard, adverse selection, and extortion of the parties amid information asymmetry. This characteristic makes the contract theory suitable for the incentive mechanism designs in FL. In the context of FL, an employer can be a model owner who wants to recruit workers to complete the FL model training. Similarly, an employee can be any client (data owner) who wants to participate in the FL. The contract theory have been widely applied in incentive mechanism design for FL. A three-dimensional contract incentive mechanism that jointly considers the task expenditure and privacy issue is design as proposed in \cite{Wu2021}. A two-period incentive mechanism based on dynamic contract is proposed in \cite{Lim2020c} to incentivize users with different willingness to participate. The contract-based personalized privacy-preserving incentive for FL in \cite{sun2021pain} can provide customized payments for workers with different privacy preferences.

        \subsubsection{Matching Theory}
        Matching theory aims to optimally match two disjoint sets of agents together, given their individual utilities. In the general allocation game model, there could be multiple agents in both sides of the matching, and agents from one side have transactions with agents in the opposite side. Therefore, such a game is called two-sided matching. In matching theory, agents compete with each other to maximize their own utility, i.e, selfishness, and always make decisions that can increase their utilities, i.e, rationality. In FL, it is used for the task allocation withe the aim of minimizing the system latency of multiple-task FL in MEC \cite{Chen2021a}, \cite{Kang2020a}.

\textbf{Summary:} This section has introduced the basis of economic and game theoretic approaches proposed to design incentive mechanisms for FL. Specifically, we provide the definitions, mechanism descriptions, and rationality behind the use of these approaches for incentive mechanisms. In the subsequent sections, we provide comprehensive reviews on the applications of economic and game theoretic approaches for incentive mechanism design in FL.

\section{Applications of Game Theory for Incentive Mechanism Design in FL}
\label{sec.application.game.theory}
In the FL service market, there are multiple participants/stakeholders which may belong to different entities, i.e., service users, FL service providers (also MOs), and data providers (also DOs). Each participant determines the optimal strategy and through constantly interacting with other agents to achieve different objectives. The objectives include the revenue, utility, cost, and system performance. The interaction among entities is complex and their objectives often conflict with each other, which makes game-theoretic approaches become effective tools for designing incentive mechanisms with low complexity for FL. With the traditional methods, it is difficult to incorporate economic implication into the solutions. Therefore, when the rationality of the stakeholders in the FL system are important, the traditional methods may not be suitable.

Game models that are commonly adopted in incentive mechanism of FL consists of Stackelberg game, non-cooperative game, and coalition game. Specifically, the Stackelberg game is used to maximize the utility of the MOs and the DOs. Otherwise, the non-cooperative game is used in the case each player, i.e., the DO or the MO, aims to maximize its own utility. To form a stable DO federation, the coalition game can be used. The game approaches are reviewed, discussed, and analyzed in the following sections.

    \subsection{Incentive Mechanisms Based on Stackelberg Game}
    In an FL system, the MO hires the DOs for model training. Thus, the MO acts as a buyer, and the DOs are sellers. The MO can first set a reward, and then the DOs decide its level of participation. To stimulate both the MO and the DOs to participate in the FL system, the Stackelberg game is adopted.

\begin{figure}[!t]
        \centering
        \includegraphics[width=3.3in]{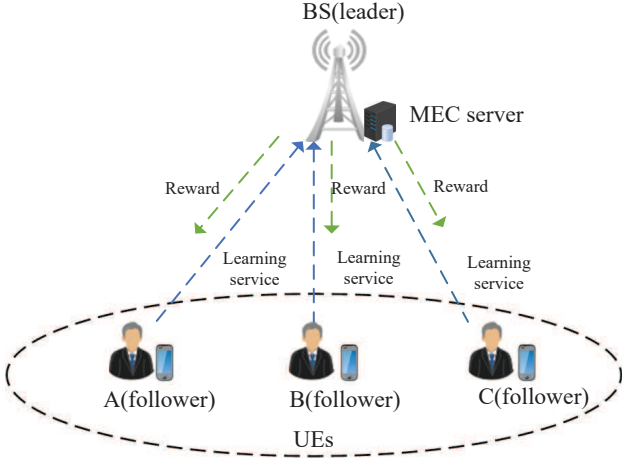}
        \caption{Stackelberg game-based incentive scheme in crowdsourcing framework.}
        \label{fig:crowdsourcing}
        \end{figure}

    The first work of Stackelberg game can be found in \cite{Khan2020}. The system model is shown in Fig.~\ref{fig:crowdsourcing}  in which the BS, i.e., the MO, is the leader, and the user equipments (UEs), i.e., the DOs, are the followers. The BS as the buyer first sets a monetary reward to maximize its utility. Given the BS's reward, each UE determines its local training strategy, i.e., the amount of CPU resources, to maximize its utility. Here, the UE's utility is a concave function of its local training accuracy and the BS's reward, and the BS's utility is a strictly concave function of the number of global iterations required to reach the global accuracy and the reward.  A unique NE among UEs can be obtained by taking the first-order condition. Given the UEs' best responses, the BS updates the global model and adjusts its reward to maximize its utility. The backward induction method is applied to solve the game. The simulation results show that an increase in the reward incentivizes the UEs to generate more local models that leads to a higher global accuracy. However, the BS needs to pay a higher incentive cost to the UEs.

    The same system model and Stackelberg game approach can also be in \cite{Pandey2019} where the MO is an MEC server, i.e., the buyer, that offers the reward. Also, the utility of the UE is the difference between the reward offered by the server and the local training cost. To obtain the best responses of the UEs, Karush-Kuhn-Tucker (KKT) and the first-order condition are used. Given the best responses, the server determines the reward to maximize its utility under the limitation of a fixed number of global iterations. Given the constraint, to achieve the desired training result, a threshold accuracy of the local training at the UEs is set. The UEs whose local accuracies are higher than a threshold are selected for the training. The threshold is optimized using the Lagrangian and Newton-Raphson method. The simulation results show that the proposed game approach outperforms the heuristic approach up to $22$\% gain in the offered reward while achieving the same target accuracy. However, the proposed game approach is constrained to a single leader, i.e., an MEC server.

    \begin{figure}[!t]
        \centering
        \includegraphics[width=2.3in,height=3in]{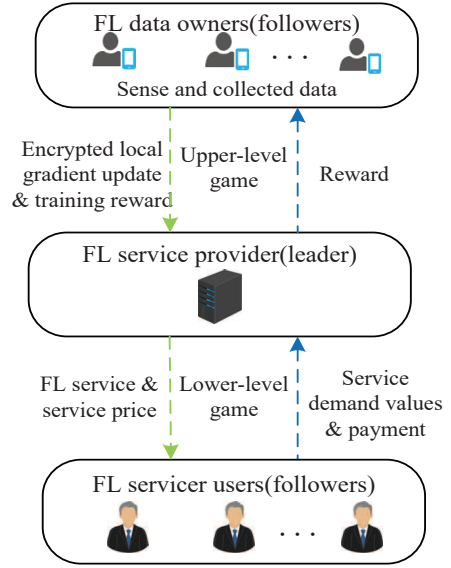}
        \caption{Market-oriented architecture for motivating information trading with federated learning based on hierarchical game.}
        \label{fig:hierarchical}
        \end{figure}

    Opposite to the model in \cite{Pandey2019}, in \cite{Lee2020}, the MEC servers of operators act as sellers (leaders), that cooperatively train a global model from a coordinator at the cloud, i.e., the buyer (the follower), based on the sensing data from IoT sensors that are distributed by the coordinator. The interaction process can be summarized as follows. First, each MEC operator individually sets its optimal price to maximize its utility, i.e, the payment from the coordinator minus its power consumption for local training, accounting its residual computational resource. Given the price and configuration profile of operators, the coordinator allocates the sensing data from IoT sensors to maximize its utility while satisfying the latency constraint of the distributed learning process. The coordinator's utility is defined as the difference between the gain owing to the model training and the payment to operators. To find an optimal data distribution strategy for the coordinator, the KKT is used. The theoretical analysis proves the existence of a unique SE for the two-level game. The simulation results show that the operators can increase the prices to improve their utilities. However, if the prices are too high, their utilities seem not to increase, since the coordinator reduces the amount of distributing input sensing data.

        \begin{table*}[!t]
        \newcommand{\tabincell}[2]{\begin{tabular}{@{}#1@{}}#2\end{tabular}}
        \centering
        \scriptsize
        \caption{Applications of Stackelberg Game for Incentive Mechanism Design in FL}
        \label{tab:table2}
        \begin{tabular}{|m{0.4cm}<{\centering}|m{1cm}<{\centering}|m{1cm}<{\centering}|m{1cm}<
        {\centering}|m{1.8cm}<{\centering}|m{1cm}<{\centering}|m{1cm}<{\centering}|m{4.5cm}
        <{\centering}|m{2.5cm}<{\centering}|}
        \hline
        \multirow{2}{*}{\bf Ref.} & \multirow{2}{*}{\bf Scenarios}
        & \multicolumn{3}{c|}{\bf Market Structure} & \multicolumn{3}{c|}{\bf Stackelberg Game} & \multirow{2}{*}{\bf Method of Solution} \\
        \cline{3-8}
        && {\bf Seller} & {\bf Buyer} & {\bf Item} & {\bf Leader} & {\bf Follower} & {\bf Objective}&\\
        \hline\hline
        \cite{Khan2020}&Edge Networks&UEs&BS&UEs' CPU resource&BS&UEs &
        Communication efficiency, global model accuracy improvement for the leader, revenue improvement for the followers & The first-order condition, backward induction method \\
        \hline
        \cite{Pandey2019}& Mobile Crowdsourcing Network& UEs&MEC server&CPU resource &MEC server&UEs &Communication efficiency, global model accuracy improvement for the leader, revenue improvement for the followers & The first-order condition, KKT condition, and backward induction method \\
        \hline
        \cite{Lee2020}&MEC Network& MEC operators &Cloud coordinator &Computational resource &MEC operators&Cloud coordinator & Revenue improvement for both leader and followers, FL latency requirement guarantee  &  KKT condition and backward induction method\\
        \hline
        \cite{Dong2020}&General FL system& SP, data workers &Service users &FL service, workers' model updates&SP&Data workers, service users& Revenue improvement for both leader and followers & The first-order derivative, backward induction method\\
        \hline
        \cite{Feng2019}&General FL system& Mobile devices &MO &Training data &Mobile devices &MO& Revenue improvement for both leaders and follower&  The first-order optimality condition, and exterior point method.\\
        \hline
         \cite{Sun2020}&Air-ground networks& Ground clients &UAVs &Local model update &UAVs&Ground clients& Reliable local model update for the leader, revenue improvement for both leader and followers &  The first-order derivative, and the second-order derivative\\
        \hline
        \cite{Hu2020}&General FL system& Mobile devices &Cloud server &FL service &Cloud server&Mobile devices& Privacy protection, revenue improvement for both leader and followers & The first-order derivative , the second-order derivative, and bisection /Newton’s method.\\
        \hline
        \cite{Chai2020}&Internet of Vehicles& Vehicles&RSUs &Training data &RSUs&Vehicles& Improved fairness, and revenue improvement for both leaders and followers &  An ADMM-based iterated algorithm\\
        \hline
        \cite{Lin2020}&Wireless charging network& WPT nodes&MEC node &Power resource &MEC node&WPT nodes& Transmission power feasibility, reward feasibility, revenue improvement for both leader and followers &  The first-order derivative, the second-order derivative\\
        \hline
        \cite{Zhan2020}\cite{Zhan2020a}&Edge network& Edge nodes&Cloud server&Traning data &Cloud server& Edge nodes& Revenue improvement for both leader and followers & The first-order derivative, the second-order derivative, and DRL\\
        \hline
        \end{tabular}
        \end{table*}

    Consider a general scenario, the authors in \cite{Dong2020} proposed to adopt a two-layer Stackelberg game to model the interactions among an FL SP (SP), users, and DOs. The system model is shown in Fig. \ref{fig:hierarchical}. In the lower-layer game, the SP is the leader, and the users are the followers. Each user determines its optimal demand value of learning service to maximize its utility. The competition among users is modeled as a non-cooperative game. The NE solution of the game is obtained by taking the first-order derivative. The utility of user is the benefit gained from the direct network effect caused by all the users in the system and the benefit gained from the indirect effect caused by other participants minus the negative impact on service quality as well as the payments from the users. In the upper-layer game, the SP acts as the leader and the DOs act as the followers. Based on the DOs' required charging price and users' service demand, the SP decides its service price to maximize its utility, i.e., its profit, which is defined as the payment obtained from users minus its fixed processing cost and rewards to the workers. Given the SP's price, each DO determines its
    charging price to maximize its utility, i.e., profit, which is the difference between the total remuneration received from SP and the additional rewards obtained owing to joining FL training and its cost for perceiving and collecting data. The first order derivative is used again to obtain the optimal strategies of the SP and the DOs. The optimal solutions of all three entities constitute the SE. However, this work does not consider constrained computing resources of DOs.

    In a practical FL network, a mobile user can be out of the coverage range of the MO. In this case, the mobile user can ask other mobile users (i.e., relay node) in the network to forward its local model to the MO. In this scenario, the authors in \cite{Feng2019} adopted the Stackelberg game in which the MO is the follower (i.e., the buyer) and the mobile users are the leaders (i.e., the sellers). The MO determines the size of dataset provided by the mobile users to maximize its utility. The utility of the MO is a function of the total data size of all the users and the prices paid to the mobile users. Given the data size, each mobile user determines the optimal price for one training data sample to maximize its profit. Here, the profit is the difference between the revenue obtained from providing the learning service to the MO and the relay service to other mobile users and the energy consumption cost. The experimental results show that the route selection of model transmission among the mobile devices can significantly improve the performance of the model and make mobile users gain the best benefit. In the future, power optimization can be considered to construct a energy-efficient cooperative FL system.

    Different from the aforementioned works, the authors in \cite{Sun2020} considered an FL network in which the MO is an unmanned aerial vehicle (UAV) as the leader, and the data owners are ground clients as the followers. The interaction between the leader and the followers is as follows. First, the UAV, i.e., the buyer, offers a reward when it publishes the FL task. Then, each client decides its participating level to maximize its utility that is the difference between the reward and the cost for training FL task. The clients' optimal strategies are determined using the first-order and the second-order derivative. Given the clients' responses, UAV determines its reward to maximize its utility. Here, the utility is defined as the total benefit gained from the training of clients minus its payment to these clients. The theoretical analysis has proved the existence and uniqueness of NE among the clients as well as the SE between the UAV and the clients. To further adapt to the dynamics of air-ground network, the authors design a dynamic incentive mechanism to adaptively select the optimal number of clients, taking into account the limitation of drone coverage. The difference between the two incentive mechanisms is the definition of the clients' loss. The simulation results show that the social welfare obtained by the dynamic incentive mechanism is higher than that obtained by the static incentive mechanism. The reason is that dynamic incentive mechanism can always select the optimal clients adapting to the time-varying environment.

    Considering the privacy leakage issue in FL, a Stackelberg game-based incentive mechanism to motivate users with sensitive data to participate in FL with guaranteed privacy is proposed in \cite{Hu2020}. The system model includes a cloud server, i.e., the MO, and mobile users, i.e., the data owners. Specifically, the server publishes an FL task and announces a total reward for attracting users. Given the server's reward, each user determines its desired privacy budget, i.e, the monetary compensation for privacy loss, to maximize its own utility. The utility of a user is defined as the payment obtained from the server minus its cost. The payment depends on not only the total reward but also the privacy budgets of users. A stable privacy budget strategy for each user can be obtained by taking the first-order and second- order derivative. Based on the optimal strategies of users, the server decides an optimal total reward that maximizes its utility, which is a function of the global model accuracy and the reward paid to the users. The optimal reward can be calculated through either the bisection method or Newton's method \cite{Ye_1994}. As shown in the numerical results, with the increasing number of users, the users' utility decreases, while the server's utility increases. However, this work only considers the users' cost associated with the privacy budget, while other costs, such as communication and computation cost, are not considered.

    To preserve the privacy of the clients, blockchain has recently been integrated with the FL as proposed in \cite{Chai2020}, which  considers an Internet of Vehicular (IoV) system. This system consists of one top chain and ground chains. In each ground chain, a multi-leader and multi-follower Stackelberg game is used to motivate vehicles to join in the FL. Specifically, the Roadside Units (RSUs) act as the leaders which are named as FL RSUs (FRs) and vehicles act as the followers which are namely FL vehicles (FVs). Each FR competes with with each other and sets price for vehicular training results, while FVs collect surrounding data for sale. Given the bid prices of all FRs and other FVs' training data sizes, one FV aims to minimize the amount of its collected data that maximizes its utility. The utility of an FV is the difference between the revenue by selling knowledge (i.e., the learning parameters of the FV's local model) to FRs according to their asking price and the computation cost for collecting and training data. Given the data size of all FVs and other FRs' pricing strategies, the goal of an FR is to minimize its bid price for purchasing FVs' knowledge while gaining more benefits. The utility of an FR is defined as the further learning reward that gains from BSs in top chain layer minus the computation cost for further learning as well as the cost for the FVs' knowledge. An iterated algorithm based on the Alternating Direction Method of Multipliers (ADMM) is adopted to deal with the proposed multi-leader and multi-player game. Simulation results demonstrate that based on the game-theoretic incentive mechanism, the proposed hierarchical FL algorithm can achieve about 10$\%$ more accuracy improvement over conventional FL algorithms. However, the training time constraints that affect the service delay of the intelligent transportation system are not considered.

\begin{figure}[!t]
        \centering
        \includegraphics[width=3.3in]{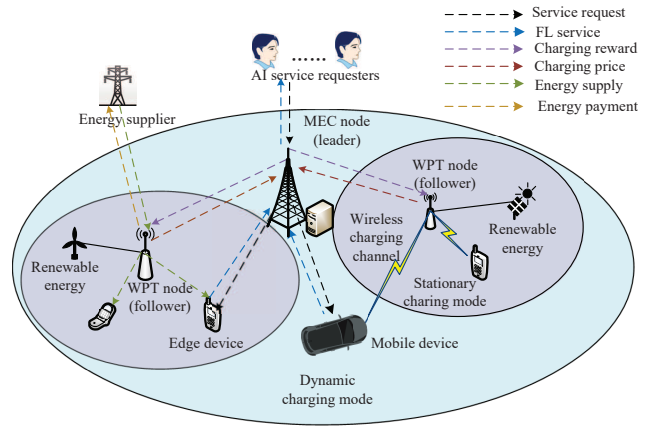}
        \caption{Energy-knowledge trading framework based on Stackelberg game.}
        \label{fig:WPEG}
        \end{figure}

    To ensure the learning performance of energy-constrained edge devices, the authors in \cite{Lin2020} proposed a novel wirelessly-powered edge intelligence framework (WPEG). As shown in Fig. \ref{fig:WPEG}, the system model consists of edge devices which provide learning services, MEC nodes which is an intermediary between the AI service requester and edge devices, and wireless power transfer (WPT) nodes which provide charging services to edge devices via wireless channel. The goal of the WPEG is to obtain the optimal power transmission and economic rewards for jointly maximizing the utility of both WPT nodes and the MEC node. The utility of WPT nodes is defined as the reward obtained from the MEC node minus the energy cost paid to the energy supplier, and the utility of the MEC node is the difference between the profit that it achieves by participating in the AI service and the reward given the WPT nodes. The optimization problem is modeled as a two-stage Stackelberg game that could be regarded as energy-knowledge trading, in which the MEC node is the leader and the WPT nodes act as followers. That is, to obtain knowledge inferred from the data of edge devices, the MEC node needs to purchase the energy from WPT nodes. Given the economic reward, each WPT node as a seller determines its power transmission strategies to optimize the self-revenue. With the WPT nodes' responses, the MEC node as the buyer adjusts its reward to maximize its utility. A low complexity gradient-based searching algorithm is designed to find the NE of the game. Numerical results show that in terms of the average utility of the MEC node, the proposed incentive scheme improves up to 32.38 $\%$, 14.42 $\%$, 41.37 $\%$, and 110.19$\%$, compared with the random federated edge learning scheme in which the MEC node offers random incentives to WPT nodes, and uniform federated edge learning scheme in which a uniform incentive is offered to WPT nodes. However, the proposed work does not consider the edge devices' mobility.

    To address the challenges of the non-publicity of individual participation strategy and ambiguous contribution evaluation, deep reinforcement learning (DRL)-based Stackelberg game can be adopted as proposed in  \cite{Zhan2020}. The system model consists of a server at the cloud that publishes an FL model and rewards, and multiple edge nodes that cooperatively train the FL model. The interaction among the server and edge nodes is modeled as a Stackelberg game where the server acts as a leader and the edge nodes act as followers. Given the server's reward and other nodes' decisions, each edge node decides an optimal training data size to maximize its utility. The utility of a edge node is defined as the reward received from the server minus its cost of data collection and model training. Given the optimal data sizes of the edge nodes, the server calculates an optimal reward to maximize its utility, which is defined as the gain of model accuracy minus the total reward paid to edge nodes. Theoretical analysis proves the existence of NE among the edge nodes and a unique SE of the entire game. Then, an actor-critic DRL model based on the proximal policy optimization is adopted to find the SE of the game based on the past strategies. The numerical results show that the DRL-based incentive mechanism can learn the optimal strategies for the server and edge nodes at which the server's utility obtained by the proposed DRL approach is higher than those obtained by the random and greedy approaches. However, how the proposed approach improves the utility of the edge nodes is not discussed.

    The same system model and hierarchical game can be found in \cite{Zhan2020a}. However, different from \cite{Zhan2020}, in \cite{Zhan2020a}, given the server’s reward, each edge nodes determines its contributed
    computational resources to the model training to maximize its utility. The utility of a edge node in each round is the difference between the payment from the server and its total CPU energy consumption for the local training to satisfy the local accuracy requirement. The optimal computational resources allocation strategy can be obtained by taking the first-order derivative. Given the best contributed resources of the edge nodes, the server optimize its rewards strategy to minimize its total cost, which is a function of the total time of one global iteration and the total payment to edge nodes. Using the KKT conditions, the best reward strategy is obtained. A DRL-based model is also designed to learn an optimal reward strategy for the server. The experiment results demonstrated comparing with greedy, random and traditional single-round optimization approaches, the DRL-based approach can achieve the best results.

    \subsection{Incentive Mechanisms Based on Non-cooperative Game}
     In the non-cooperative game, all the players simultaneously decide their strategies, while in the Stackelberg game, one or some players as leaders decide their strategies before other players (followers).

\begin{figure}[!t]
        \centering
        \includegraphics[width=2.7in, height=2.8in]{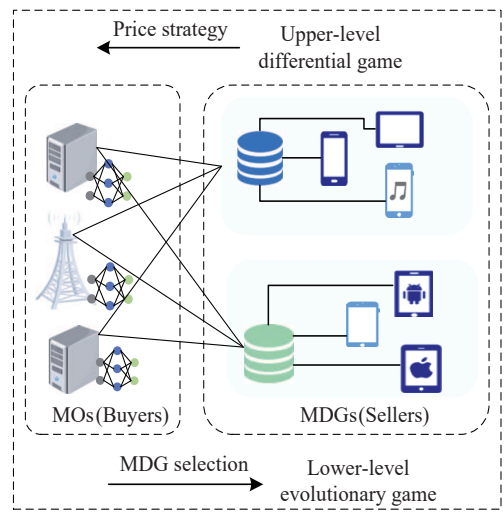}
        \caption{Federated learning training service market based on dynamic game.}
        \label{fig:MDG}
        \end{figure}

    In \cite{Zou2019} and \cite{cheng2021dynamic}, the authors considered a federated learning training service market. The system model is shown in Fig. \ref{fig:MDG} which consists of multiple MOs and multiple mobile device groups (MDGs). The MOs as buyers select MDGs, i.e., the sellers that have access to a federation of enormous mobile devices, to complete the FL training service. A two-level dynamic game is adopted to formulate the interactions in this training service market. In particular, the MDG selection are modeled as a lower-level evolutionary game, while the pricing strategies of MDGs are formulated as a higher-level differential game. Given the MDGs' price, each MO adjusts its selections according to the price and time-varying observed model accuracy to maximize their utility. The utility of MO is defined as a net profit function, i.e., the gain of accuracy from the MDG minus the price paid for the training service and congestion punishment from MDG. Given the MOs' selections, each MDG optimizes its price as the best response to MOs' selections and other MDGs' pricing strategies so as to maximize their accumulative profits gained from the training services. The profit of the MDG is defined as the difference between the payment for selling data and the cost for training. The differential game can be regarded as an optimal control problem which can be solved by using the iterative numerical algorithm \cite{tabak1975numerical}. The simulation results show that the accumulative profits of MDGs obtained by the differential game approach are much higher than those obtained by the static non-cooperative game.

    Considering the same FL system as described in \cite{Zou2019}, the evolutionary game model \cite{Zou2019a} is formulated to analyze the dynamic training strategies of the mobile devices with bounded rationality. The competitive evolution of individuals (mobile users) in the population is modeled. However, unlike \cite{Zou2019}, the utility, i.e., the payoff, that one mobile device can receive from the MO is proportional to the data size that the mobile device uses for the training. The objective of each mobile device is to maximize its individual utility, and to achieve it, the mobile device dynamically adapts its strategy according to its utility. In particular, each mobile device compares its utility with the average utility, and if the utility is less than the average, it will select another strategy in the next period. The uniqueness and stability of the evolutionary equilibrium is proved by using the Cauchy-Lipschitz theorem \cite{DiPerna1989} and Lyapunov’s second method for stability \cite{sastry1999lyapunov}, which allows the mobile users to have the optimal strategies. Numerical results show that the utilities fluctuate first over time and then converge to the evolutionary equilibrium and the learning rate decreases dramatically over time, which is consistent with the theoretical analysis.

\begin{figure}[!t]
        \centering
        \includegraphics[width=3.3in]{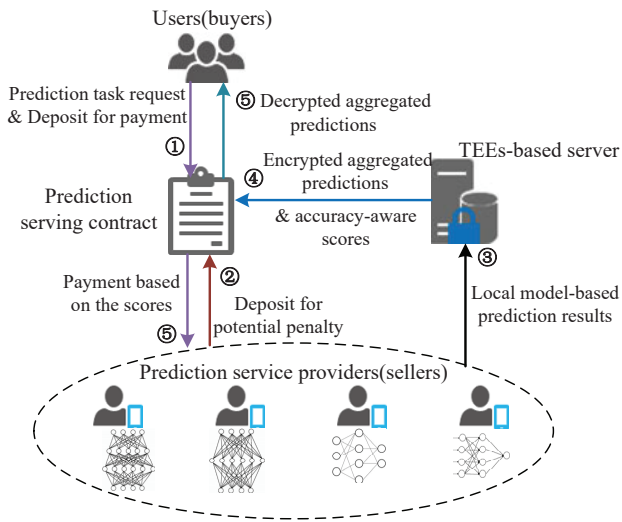}
        \caption{Federated prediction serving framework based on Bayesian game.}
        \label{fig:prediction}
        \end{figure}

    To motivate the self-interested MOs to provide truthful results for prediction services, a non-cooperative Bayesian game-based incentive mechanism is proposed in \cite{Weng2020}. The system model is shown in Fig. \ref{fig:prediction} which consists of prediction SPs, users, server and a smart contract. Specifically, the SPs who own various ML models monetize their prediction query services on the blockchain. The user can query the prediction services through the smart contract. Moreover, the contract will assign the users' deposit to the providers according to the respective accuracy-aware scores calculated by the server. In addition, if the SP does not offer the promised prediction, it will be punished through submitting a deposit. The strategy of the SP is to choose whether to participate in a prediction service, and if participating, what quality level local model to provide. The utility is the difference between the payment from users and the cost for providing prediction and its deposit. To obtain the optimal behavior strategies of SPs, the objective of the proposed mechanism is to maximize the overall probability of Kullback-Leivler divergence between each SP's prediction and the true prediction, subject to IR and BB. A Bayesian NE can be obtained through applying the  divergence-based Bayesian Truth Serum method \cite{Radanovic2014}. The simulation results show that the participating SP who offers a truthful prediction will receive sufficient incentives which demonstrates the effectiveness of the proposed pricing mechanism. However, the scenario where the input data is injected into the imperceptible perturbations is not considered.

    To detect and discard bad updates provided by clients, the authors in \cite{Tahanian2021} model the aggregating process in the FL system by a mixed-strategy game. The players are a server and the clients. The valid strategies of the clients are to send good or bad updates while the server can accept or ignore these updates. When a client sends good updates and the server accepts them, both of them will earn some payoffs. A trustworthy metric is used to measure a good or bad update. This metric is inversely proportional to the distance between the local model transmitted by the client and the aggregation model. Then, the NE property is applied to determine the probability of different strategies of the server and clients. At the mixed Nash equilibrium, both the players have no an incentive to change their strategies since they are not able to increase their expected payoff by selecting an alternate strategy. The simulation results show the proposed algorithm can detect 100$\%$ of the bad clients, and the test accuracy of the proposed algorithm is at least 15.8$\%$ and 2.3$\%$ better than the previous works (e.g., Multi-KRUM \cite{BlanchardMGS17}, Federated Averaging \cite{McMahanMRHA17}, and COMED \cite{KonecnyMYRSB16}) for flipping and noisy scenarios, respectively. The detection of the backdoor attack where a malicious client can use model replacement to introduce backdoor functionality into the global model can be considered in the future work.

\begin{figure}[!t]
        \centering
        \includegraphics[width=3.3in]{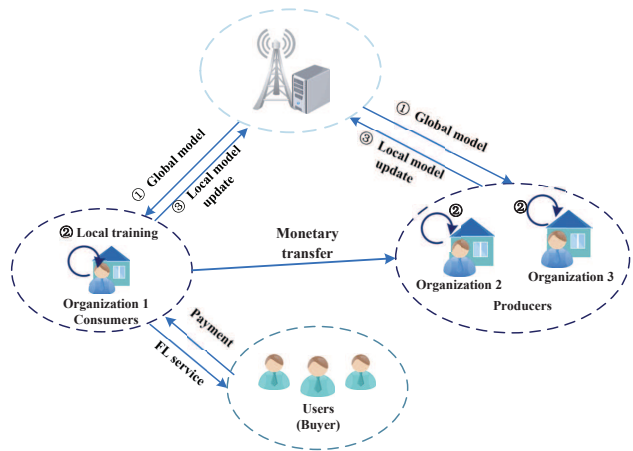}
        \caption{The trading market based on non-cooperative game in cross-silo FL system. Note that the diagram only shows the scenario that organization 1 acts as the consumer which uses the model trained by all three organizations.}
        \label{fig:cross-silo}
        \end{figure}

    Considering the organization heterogeneity and non-excludable public goods (i.e., resources of organizations), the authors in \cite{Tang2021} proposed an incentive mechanism for cross-silo FL. As shown in Fig. \ref{fig:cross-silo}, in cross-silo FL, each organization is not only the data owner that cooperatively trains a global model from a third party entity, but also the owner of the global model that utilizes the resources of organizations to gain revenue from its market. In the proposed incentive scheme, each organization submits the message files, including the number of training rounds that it can participate in and the unit monetary reward per training round that it hopes to gain, to the server. Once receiving organizations' message profile, the server calculates and announces the processing capacity used by each organization for local training and their respective monetary transfer received from or paid to other organizations. Each organization optimizes its message profile to maximize its payoff, which is defined as its utility and additional monetary transfer minus the cost. The utility of the organization is a function of the difference between the precision of the global model without training and that of the trained global model. A non-cooperative game with perfect information is used to model the strategic interaction among organizations, which is formulated as a social welfare maximization problem, i.e., maximizing the overall payoffs of all organizations. In order to obtain the NE solution, a distributed algorithm inspired by the distributed accelerated augmented Lagrangian method \cite{Nikolaos2015} is proposed. The simulation results show that the proposed mechanism can enable the organizations to achieve a higher social welfare through participating in cross-silo FL. A scenario that the organizations determine the numbers of participating rounds considering their valuation on precision and their computational and communication costs can be considered in the future.

        \begin{table*}[!t]
        \newcommand{\tabincell}[2]{\begin{tabular}{@{}#1@{}}#2\end{tabular}}
        \centering
        \scriptsize
        \caption{Applications of Non-cooperative Game and Other Games for Incentive Mechanism Design in FL}
        \label{tab:table3}
        \begin{tabular}{|m{0.3cm}<{\centering}|m{1.1cm}<{\centering}|m{1.3cm}<{\centering}|m{1.1cm}
        <{\centering}|m{1.1cm}<{\centering}|m{1.2cm}<{\centering}|m{4.6cm}<{\centering}|m{2.5cm}
        <{\centering}|m{1.3cm}<{\centering}|}
        \hline
        \multirow{2}{*}{\bf Ref.} & \multirow{2}{*}{\bf Scenario}& \multirow{2}{*}{\bf Game Model} & \multicolumn{3}{c|}{\bf Market Structure} & \multirow{2}{*}{\bf Mechanism}& \multirow{2}{*}{\bf Objective}& \multirow{2}{*}{\bf Solution} \\
        \cline{4-6}
        &&& {\bf Seller} & {\bf Buyer} & {\bf Item} &&&\\
        \hline \hline
        \cite{Zou2019} \cite{cheng2021dynamic}&General FL system&Evolutionary game, differential game&MDGs&MOs&MDGs' learning service&
        Given the training price offered by the MDGs, the evolutionary equilibrium MDG selection strategy for the MOs is determined.&Individual utility maximization for buyers, and the accumulative profits maximization for sellers&Hamilton solution, evolutionary equilibrium \\
        \hline
        \cite{Zou2019a} &General FL system&Evolutionary game&Mobile devices&MOs&Mobile devices' data&The training strategies selection problem of the mobile devices is formulated as evolutionary game. The uniqueness and stability of the evolutionary equilibrium solution is proved.& Sellers' individual utility maximization&Evolutionary equilibrium\\
        \hline
        \cite{Weng2020}&FL prediction serving framework&Non-cooperative Bayesian game&Prediction SPs&Users&The providers' prediction service&The strategic behaviors of participating providers in federated prediction is modeled as a Bayesian game. The Bayesian Nash equilibrium participation strategy is obtained applying the idea of divergence-based BTS method.&Truthfulness, individual rationality, budget feasibility, fair reward, and sellers'  utility maximization&Bayesian Nash equilibrium\\
        \hline
        \cite{Tahanian2021}&Robust FL&Mix-strategy game&Clients&Central server&Clients' learning service&Sellers provide good or bad updates for the buyer, and the buyer selects to accept or reject these updates. The probability that the clients provide good updates is calculated by using mix-strategy game.& Payoff maximization for both the sellers and the buyer.&Mixed strategy Nash equilibrium\\
        \hline
        \cite{Tang2021}&Cross-silo FL&Non-cooperative game&Organizations&Organizations&Organizations' learning service&Once receiving the participated round and monetary transfer from the organization, the central serve calculates them. The sellers determine their optimal decisions through solving the social welfare maximization problem using the Lagrangian.& Efficiency, individual rationality, budget balance, and sellers' payoff maximization& Nash equilibrium\\
        \hline
        \cite{Hu2020a}&FEL&Cooperative game&Edge devices&Edge server&Edge devices' learning service&The equilibrium solution for the devices' participating strategy is derived through decomposing the original problem and solving it.& Global profit maximization for the buyer, and guarantee incentive compatibility for the sellers&Equilibrium    solution\\
        \hline
        \cite{Hasan2021}&General FL system&Hedonic game&Data owner agents&MAE&Agents' learning service&Agents selects an optimum cluster that enable their utility to be improved to join, and the utility of each players is allocated by MAE according the marginal contribution.& Stable cluster formation and budget balance& Nash equilibrium\\
        \hline
        \cite{Lim2020}&FL in mobile networks&Contract theory, coalition game&Workers&Model owners&Workers' training data& The sellers choose optimum contract bundles for utility maximization and contribute the corresponding data to the training. The buyers adjust their reward strategies and collaborative with other owners to form federation to maximize their payoffs.&Guarantee individual rationality, incentive compatibility and revenue improvement for workers, buyers' marginal contribution maximization&Optimal solution\\
        \hline
        \cite{Qu2021}&Proof of FL&Reverse game&Data providers&Pool members&Data providers' training data&Given the sellers' data pricing, and then the buyers determine their optimum bid. The Euler-Lagrange equation and variational method are used to obtained the best strategies of both parties.&The utility maximization for both buyers and sellers&Optimal solution\\
        \hline
        \end{tabular}
    \end{table*}


    \subsection{Incentive Mechanisms Based on Other Game Methods}
    To stimulate the users to participate in the FL system in a long-term, the fairness allocated to the users should be guaranteed. To achieve fairness, hedonic game and coalitional game are appropriate solutions. Similar to \cite{Zhan2020}, the authors in \cite{Hu2020a} considered an FEL system in which edge devices are not only producers but also consumers. Differently, they studied the utility maximization problem only for edge devices. This relationship is modeled as a participation game, and each edge device acts as a player. In the game, each edge device makes a decision on whether to join a round of FEL so as to gain a payoff, given other devices' participation decisions. The payoff of the device is the difference between the income owing to serving the device users and the cost consumed for local training and updating. The device's objective is to maximize its payoff with the optimal participation decision. Considering the privacy among edge devices, a global profit maximization (GPM) problem, i.e, maximizing the overall expected profits of all devices, is formulated to find the best participation strategy for the device from the global viewpoint. And the definition of correlated equilibrium is introduced to guarantee that no player can deviate from the assigned strategy to maximize its profit given others' strategies. To solve the optimization problem in polynomial time, the original problem is decomposed into several sub-GPM problems according to the number of edge devices. The simulation results show that the device who has larger data contribution than others is more likely to participate in the FEL. However, the learning service latency is not taken into account which is important to the edge intelligence service.

    To deal with stable clustering problem of data owner agents contributing to federated training process, in \cite{Hasan2021}, the hedonic game is used to simulate the problem where the agents act as the players of the game and the clusters as the coalitions. The system model includes multiple agents (i.e., sellers) which cooperate with other agents selectively to form clusters and train model, and a model aggregating entity which as a utility allocator (i.e., buyers) is responsible for assigning a utility to all possible clusters. The utility of any cluster is the sum of the raw utility and learning utility. The strategy of the agent is to decide which cluster to join with the aim of maximizing its utility, and the utility is the sum of the minimum payment price that any agent can obtain if it join a cluster and the gain owing to joining a specific cluster. To ensure the Nash-stability of the hedonic game, the definition of the Nash-stable set is introduced to design an effective utility allocation method for agents. For all clusters, the utility allocation method satisfies the property of budget balance since the overall utility of all agent in one cluster is not greater than the cluster's utility. The theoretical analysis prove that when the utility function of a cluster is superadditive and preferences of a play are additively separable and symmetric, it always admit a Nash-stable coalition partition.


    Unlike \cite{Hasan2021 } and \cite{Donahue2020} which studied the federation among DOs, the federation of multiple MOs is considered in \cite{Lim2020}. Firstly, to motivate the worker to participate in FL, a contract theory-based incentive mechanism in mobile crowdsensing network is proposed. In this system, the MOs (buyers) train local models based on the data collected by workers (sellers), and collaborate with other MOs selectively to form federations to maximize their own profits. Given the prevailing federation parameters (i.e., others' strategies for workers' contribution), every MO's contract-theoretic incentive mechanism is designed separately with the aim of maximizing their marginal contribution while minimizing incentive cost paid to workers. The strategy of the MO is to determine an optimal reward to maximize its profit, which is defined as its marginal contribution minus the incentive cost and resources cost for model training. The marginal contribution of the MO is the utility it produces after joining the FL. Moreover, the strategy of each worker is to determine its data quantity contribution to maximize its individual utility, i.e., the difference between the reward from the MO and the cost incurred by collecting data. Given a set of feasible data contribution subsequences, the optimal reward strategy is obtained by using the approach in \cite{Nguyen2016}. Using the optimization tool, i.e., cvxpy \cite{Diamond2016} or the Bunching and Ironing algorithm \cite{Gao2011}, an optimal set of data contribution is obtained. Then, a coalitional game is designed to model the problem of federation formation, and the merge and split algorithm proposed in \cite{Apt2006} is adopted to study the stable federation partition given different characteristics of data quantity and quality the MOs have. The numerical results show that the reward of workers increases as its data contribution increases and there exists the federation formation equilibrium in the proposed mechanism design. However, there may be competing or malicious MOs which can affect the performance of a federation adversely. Furthermore, the reputation mechanisms \cite{Jaramillo2010} can be introduced to choose reliable MOs.

    In \cite{Qu2021}, a novel consensus mechanism, proof of FL, is proposed, in which the pool members are the buyers to cooperatively train a global model from the pool manager based on their privacy data purchased from the data providers (i.e., sellers). To guard against the risk of privacy data leakage in FL, a reverse game-based data trading mechanism is proposed which enables a pool member (miner) to maximize its utility only when it trains the model without any leakage. The strategy of the data provider is to determine an optimal data pricing rules to maximize its utility, which is the final price of the traded data minus the expected loss brought by sensitive data leakage. Given the data provider's pricing rules, the pool member calculates its optimal bid according to its real privacy information, i.e, the IC property, for maximizing its utility. The utility is the legally expected net profit from this data trading and the expected profit of leaking sensitive data minus the final price of the traded data. The optimal strategies of both parties can be obtained through solving the Euler-Lagrange equation of their utility by adopting the variational method. The simulation results show that the higher reputation value of the pools , the bigger probability of successful data trading and the higher utility of both parties.

\textbf{Summary:} In this section, we have reviewed the applications of game theory approaches for the incentive  mechanism design in FL. The objective is to find the equilibrium solution among players to achieve utility maximization, while satisfying IR, IC and BB. The Stackelberg game-based approaches are summarized along with the references in Table \ref{tab:table2}, and Table \ref{tab:table3} summarizes the approaches based on non-cooperative game and other games. As observed from the tables, the Stackelberg game are used to maximize the utilities for both the MO and the DO while non-cooperative game is mostly adopted to maximize the utility of DOs with competitive relationship. In fact, auction is also an effective incentive method to motivate the players to join in the FL, especially in a high competition environment with limited resources. The next section discusses how to apply auctions for incentive mechanism design in FL.

\section{Applications of Auction for Incentive Mechanism Design in FL}  
\label{sec.application.auction.fl}
Auctions have some good properties. For example, IC encourages the clients to report true cost that gain the utility of the model owner or encourages the model owner to report its true value of the computing resource, while IR guarantees that every client joining the FL has a non-negative utility. Thus, auction is a very effective solution for the incentive mechanism in FL.

    \subsection{Incentive Mechanisms Based on Sealed-bid Auctions}
    In a sealed-bid auction, sealed bids from bidders are simultaneously sent to the auctioneer. This approach is mainly applied in participant selections for FL. The MO sends a request of FL task to the DOs. After receiving the service request, each DO (as a bidder) submits a message to the MO (as the buyer/auctioneer), which includes data such as bidding information, cost, and computation capacity. Particularly, the bidding information submitted by each DO is not open in the auction and FL trading process. In each round of auction, the MO selects the DO which meets the training conditions of FL at this stage through solving a winner selection problem. The selection process is repeated until the FL task can be completed.
        \subsubsection{FPSB Auction}
\begin{figure}[!t]
                \centering
                \includegraphics[width=2.4in,height=2.7in]{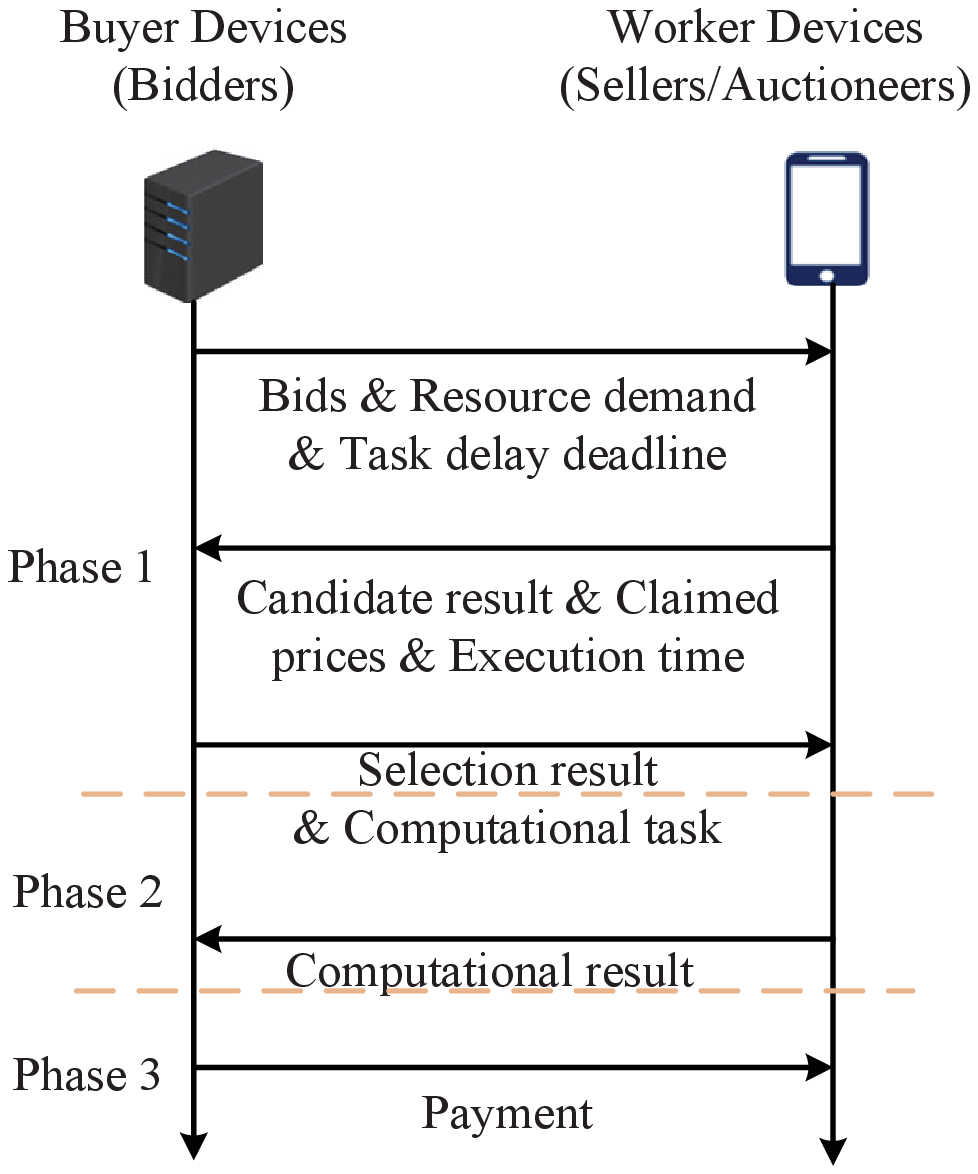}
                \caption{Double auction based distributed task allocation in mobile device cloud.}
                \label{fig:double}
            \end{figure}

        To motivate data owners to contribute their idle resources for the performance improvement of computation-intensive application, the authors in \cite{Roy2021} designed a double auction-based \underline{FE}derated learning and \underline{S}ubjective logic-driven distributed \underline{T}ask allocation system in mobile device cloud, namely FEST. In the FL market, it is necessary to match the bidders and the sellers since there are multiple bidders (model owners) and sellers (data owners). In FEST, buyer devices as bidders recruit high-quality worker devices, i.e., sellers, with reliable and experienced execution performance to achieve the best system performance. The FEST-auction process consists of three phases: bidder selection, task allocation and worker payment assignment, as shown in Fig.~\ref{fig:double}.

        In the first phase, after receiving the task requests (including buyers' task bid, resource requirement and delay requirement of the task) from the bidder, each seller computes the priority factor of each bidder, which is an integrated metric with related to task bid, execution time and preference to seller. Then each seller greedily selects bidders sorted based on the descending order of priority factor until it satisfies the resource, energy and maximum workload constraints. In the second phase, a bidder only recruits one seller for a computing task. In fact, a bidder can win bids from multiple sellers. Therefore, each bidder chooses the best one from multiple candidate sellers which can maximize its utility. The utility of the bidder is a composite function of the bidder's valuation for resources and the seller's ask price for resources as well as its execution time and reputation value. In addition, a payment policy based on the approach in \cite{Ren2015} is designed where the seller providing high-quality service is rewarded whereas the seller failing to rendering the required task quality is punished. Theoretical analysis proved that the proposed FEST framework achieves the properties of CC, IR, truthfulness, and BB. Simulation results imply that the FEST system achieve performance improvement as high as 30$\%$ and 25$\%$ in terms of quality-of-experience and utility of the bidder, respectively, compared with the truthful incentive mechanism (TIM) \cite{Jin2015} and distributed truthful auction mechanism (DTAM) \cite{Wang2021} schemes. FEST performs better than the two existing approaches since they only consider the resource availability of sellers while not considering the quality-of-experience of the bidders. However, the growth of the number of FL worker will lead to much higher computational complexity, which makes the FEST based auction approach not suitable to large-scale FL applications.

        \subsubsection{SPSB Auction}
        The aforementioned works are for resources allocation in which only a single FL service is considered. However, as the diversification of ML-driven applications develop and grow, it is predicted that the wireless network will host multiple co-existing FL services. To fill the gap, the authors in \cite{Xu2021} considered a multi-FL service scenario, where FL service providers (SPs) cooperate with each other to maximize the overall system performance or are selfish to maximize their own performance. For the cooperative FL SPs, intra-service and inter-service resource allocations are considered. The optimal FL round length  is obtained in intra-service problem through optimizing the bandwidth allocation among the clients of each FL service. The optimal solution to the intra-service bandwidth problem is then used to solve the inter-service problem, and the bandwidth allocated to each FL SPs is computed by using dual decomposition \cite{Palomar2006}. In addition, to address the selfishness issue among FL SPs, a multi-bid auction mechanism is designed to allocate bandwidth to SPs so as to maximize the SPs' utilities. The network operator (seller) finds an optimal allocation scheme to maximize each SP's (buyer) utility through the use of the KKT condition. Each SP is then charged according to the second-price auction payment policy. The simulation results show that the proposed two bandwidth allocation schemes are superior to benchmark algorithms in terms of inherent fairness by design, where bandwidth is equally or proportionally allocated  among clients and SPs. However, when a client can simultaneously participate in multiple FL service, resource allocation should take into account both bandwidth resource and client computational resources.

        Given the growing complexity and capacity of the future network, it is difficult to completely model or solve the dynamic conditions in the network with the standard approaches. The authors in \cite{Lim2020a} proposed deep learning-based (DL) optimal auction incentive scheme, where multiple MOs (buyers) intend to buy the 3C-L resources (i.e, communication, computation, caching, and learning resources) from a worker (seller). In this incentive mechanism, the concept of information freshness, i.e., Age of Information (AoI) is adopted. Meanwhile, the DL-based pricing scheme is proposed to price the workers contribution in terms of AoI, in which both conditions of truthfulness and revenue maximization of the seller are ensured. Although the proposed incentive scheme uses the second-price auction to determine the winner, it allows the worker to earn higher revenue than that of the conventional second-price auction scheme. The reason is the information asymmetry among the buyer and the sellers can affect the performance of the latter scheme, but the former can avoid this issue through learning from the historical data so as to improve the performance.

        To overcome the issue of the failure of the communication link and missing nodes that may affect the performance of FL, a communication-efficient FL scenario where the UAVs as wireless relays to improve the communication between IoV components (workers, i.e., DOs) and the FL server is considered as proposed in \cite{Ng2020a} and \cite{Ng2021}. The authors presented a joint auction-coalition formation framework to allocate the UAVs coalitions to cells of workers. Note that the payment that a cell receives depends on the total time needed to complete the FL task. Therefore, the workers in a cell have higher incentive to bid for the UAV coalitions that can maximize the utility of the cell(i.e., the difference between the valuation of the cell for a coalition and its payment price), and then and pay them based on the second-price auction. Furthermore, the UAVs aim to maximize sum of their individual profits through finding a stable partition. To achieve the objective of both cells and UAVs, a merge-and-split algorithm is proposed to find the optimal coalitional structure. Simulation results show that the proposed join auction-coalition scheme achieves the highest profit of coalition, compared with the merge-and-split with random allocation and random partitioning with second-price auction. In the future, a joint trajectory optimization and energy efficient coalition formation game can be considered to complete the FL tasks more efficiently.

        \subsubsection{VCG Auction}
        In \cite{Cong2020a}, the authors proposed Fair-VCG (FVCG) as an improvement of VCG mechanism for incentivizing DOs to contribute their data and truthfully report their cost in FL training processes. FVCG aims not only to maximize the social welfare but also minimize unfairness of the learning federation without the FL server (buyer) knowing the cost type and data quality of DOs (sellers). Guaranteeing the fairness among DOs can encourage them to improve their participation level. The proposed FVCG approach mainly includes two steps: 1) the calculation of the data acceptance vector, which illustrates the FL server decides how much data to accept from each DO; 2) the calculation of the payment vector (including the VCG payment vector and the adjustment payment vector), which indicates the payment each DO can receive from the server. In particular, the adjustment vector is defined as the solution of a functional optimization problem, which is difficult to solve. FVCG adopts an unsupervised and composite neural network based approach to solve the optimization problem since the neural network can approximate continuous function with an arbitrary precision. The experimental evaluation results reveal the reasonableness of FVCG and the proposed neural network-based approach.

        The proposed incentive mechanism in \cite{Cong2020a}, i.e., FVCG, overcomes the supply-side information asymmetry and ensures the fairness among the DOs in the FL process. In practice, the information asymmetry is multidimensional including the demand side and the supply side, both of which affect the sustainability of federation. For this reason, a game-theoretical model for cooperative production of virtual products is presented in \cite{Cong2020} and \cite{Cong2020b}. Specifically, a mechanism similar to FVCG, Procurement-VCG (PVCG) is proposed on the supply side, and the Cr\'{e}mer-Mclean mechanism \cite{Cremer1985} is used to overcome the demand-side information asymmetry (the model users’ valuations on the trained FL model). Differently, in PVCG, if one producer (i.e., the seller) cannot offer the data it claims at the time of bidding, the coordinator (the intermediary between producers and model users, i.e., the buyer) will impose a high punishment on it. The theoretical analysis demonstrated that the Cr\'{e}mer-Mclean mechanism, together with the procurement auction, maximizes producer surplus by motivating producers to offer all their data to the federation and truthfully report their private information. However, when the number of producers increase, more advanced algorithms is required for FVCG and PVCG to learn the two payments. In addition, the effectiveness and superiority of FVCG and PVCG need to be verified compared with other sharing rule (e.g., SV, \cite{Jia2019, Ohrimenko2019} and Labour Union \cite{Gollapudi2017}).

    \subsection{Incentive Mechanisms Based on Reverse Auctions}
\begin{figure}[!t]
            \centering
            \includegraphics[width=2.4in,height=2.7in]{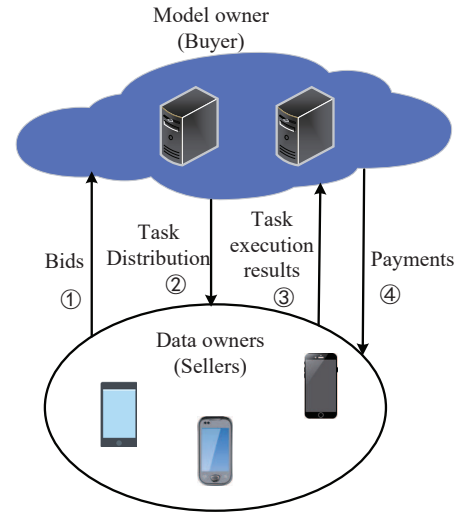}
            \caption{Reverse auction based incentive mechanism for client selection in FL.}
            \label{fig:reverse}
        \end{figure}

     For reverse auction, sellers bid for the prices at which they are willing to sell their resources. One of major advantages of the reverse auction is to incentivize the participants, e.g., the data owners, to report their types, e.g., the cost of FL training. This advantage is important to the model owner since it helps to reduce the incentive cost. As shown in Fig. \ref{fig:reverse}, there are four steps in a typical reverse auction based incentive mechanism including bidding, winner selection and task assignment, feedback of the task results, and payment determination.

    In \cite{Le2020}, the authors modeled the incentive mechanism between the BS and multiple mobile users as a reverse combinatorial auction game, where the BS acts as the buyer (also the auctioneer) and the mobile users as the sellers are allowed to bid for the combination of resources. In particular, each mobile user sends a message consisting of the required amount of resources, local accuracy, and the corresponding energy cost to the BS. Given the maximum tolerable time of FL (the upper bound of the total time of one global iteration), each mobile user determines the optimal bid, which includes the uplink transmission power, the CPU cycle frequency and the local accuracy level, to minimize its energy cost. And a low-complexity iterative algorithm is proposed to obtain the optimal solution of the problem. Once receiving the bids from all mobile users, the BS performs winner selection and payment determination by solving an social welfare maximization (SWM) problem using the primal-dual based greedy algorithm. Note that the utility of BS is the difference between the BS's satisfaction level (based on the local accuracy that the user can offer in the bid) and the payment for mobile users. Simulation results show that the social welfare obtained by the proposed mechanism is 400$\%$ larger than by the fixed price scheme since the fixed price mechanism heavily depends on the prices of the resources.

    To avoid the collusion between the untrustworthy third party and the ENs, the authors in \cite{Fan2021} presented a a decentralized and transparent blockchain-based resource
    trading system for FL in edge computing (EC). To encourage both the resource requester (individuals or companies
    who need model training) and ENs (the owner of data and computing capacity) to join in the system, the incentive problem among them is formulated as a data quality-driven reverse auction (DQDRA) with the aim of maximizing the valuation of the requester about the model accuracy. Accordingly, ENs, i.e, sellers, determine the optimal bids which can maximize their utilities (including data size, the EMD metrics, i.e., a metric for data distribution, and prices) and submit them to the requester, i.e, the only one buyer. Once receiving the ENs' information, the requester decides the winner and the payment to them based on the proposed monotone greedy algorithm by marginal density for its valuation maximization. The service valuation of the requester is a function of training data size. The simulation results show that the valuation of the requester in DQDRA is larger than the valuation of the requester in the GREEDY-SM and RANDOM-SM mechanisms proposed in \cite{Chen2011}. The trading market with multiple competitive requesters can be taken into account in the future.

    The same system model and auction mechanism to that in \cite{Fan2021} can be also found in \cite{Jiao2019}. However, in \cite{Jiao2019}, the buyer is an FL platform, and the sellers are DOs, i.e., FL workers, which report the requested wireless channel to communicate with the platform. The FL training problem between the platform and DOs is converted to a SWM problem where
    the utility of the platform is the data utility minus the total cost and the total payments to workers.
    To tackle the SWM problem, the original auction is decomposed into a set of sub-auction in terms of EMD. After that, each group greedily selects the workers and determines the payments based on the marginal virtual social welfare density. To further improve the social welfare, the author proposed a DRL-based auction (DRLA) which use the graph neural network to extract the features from workers' reported types and automatically decides the service allocation and payment. The experimental results show that the DRLA and RMA can achieve higher social welfare compared with the benchmark scheme only depending on the bidders' bid prices, and the social welfare is improved in DRLA.

    \begin{table*}[!t]
    \label{table.4.app.sealed.fl}
        \newcommand{\tabincell}[2]{\begin{tabular}{@{}#1@{}}#2\end{tabular}}
        \centering
        \scriptsize
        \caption{Applications of Sealed-Bid Auction Model for Incentive Mechanism Design in FL}
        \label{tab:table4}
        \begin{tabular}{|m{0.4cm}<{\centering}|m{1.6cm}<{\centering}|m{1.1cm}
        <{\centering}|m{1.1cm}<{\centering}|m{1.2cm}<{\centering}|m{5.7cm}
        <{\centering}|m{3.7cm}<{\centering}|}
        \hline
        \multirow{2}{*}{\bf Ref.} & \multirow{2}{*}{\bf Auction Model} & \multicolumn{3}{c|}{\bf Market Structure} & \multirow{2}{*}{\bf Mechanism}& \multirow{2}{*}{\bf Objective}\\
        \cline{3-5}
        && {\bf Seller} & {\bf Buyer} & {\bf Item} &&\\
        \hline\hline
        \cite{Roy2021}&Sealed-bid double auction&Worker devices&Buyer devices&Computational resources&Based on buyers' bids, the sellers determines buyer winning candidates by using the greedy approach. The seller that can maximize the utility of the buyer is selected.&Execution time of FL task minimization, Buyer devices' utility maximization, CC, IR, BB, and truthfulness\\
        \hline
        \cite{Xu2021}&Second-price sealed-bid auction&Network operator&FL service providers&Bandwidth&Buyers submit to the auctioneer their bids including the requested bandwidth and the unit price, the auctioneer allocates bandwidth to buyers based on the market cleaning price.& Utility maximization for SPs, IR, IC and fairness\\
        \hline
        \cite{Lim2020a} &deep learning-based optimal auction&Worker&MOs&3C-L resources& MOs' bids are transformed to new bids satisfying IR and IC properties, the winners are decided based on second-price auction and the payment price of them is derived using ReLU.&Revenue maximization for the seller, IR, and IC, best matching\\
        \hline
        \cite{Ng2020a} \cite{Ng2021}&Second-price sealed-bid auction&UAVs&Workers&Train resources& Buyers submit their bids to UAV coalition with sufficient energy, and then each coalition is allocated to cells of workers.&Total profit maximization for sellers, IC,IR\\
        \hline
        \cite{Cong2020a}&VCG auction&DOs&FL server&Training data&Based on the data quality and privacy cost of sellers, the buyer determine the amount of acceptable data to maximize its revenue. The payment function is derived by using composite neural networks.& IR,IC, SWM, WBB and fairness\\
        \hline
        \cite{Cong2020}&VCG auction&Producers&Consumers&Virtual products&Sellers submit to the auctioneer their bids including their capacity and type parameters, the coordinator determines the acceptance ratio and transfer payment to sellers.&SWM, truthfulness, ex-post allocative efficiency, ex-post IR, WBB\\
        \hline
        \end{tabular}
    \end{table*}

    \begin{table*}[!t]
    \label{table.5.app.reverse.fl}
        \newcommand{\tabincell}[2]{\begin{tabular}{@{}#1@{}}#2\end{tabular}}
        \centering
        \scriptsize
        \caption{Applications of Reverse Auction Model for Incentive Mechanism Design in FL}
        \label{tab:table5}
        \begin{tabular}{|m{0.4cm}<{\centering}|m{1.6cm}<{\centering}|m{1.1cm}
        <{\centering}|m{1.1cm}<{\centering}|m{1.4cm}<{\centering}|m{5.5cm}
        <{\centering}|m{3.7cm}<{\centering}|}
        \hline
        \multirow{2}{*}{\bf Ref.} & \multirow{2}{*}{\bf Scenario} & \multicolumn{3}{c|}{\bf Market Structure} & \multirow{2}{*}{\bf Mechanism}& \multirow{2}{*}{\bf Objective}\\
        \cline{3-5}
        && {\bf Seller} & {\bf Buyer} & {\bf Item} &&\\
        \hline\hline
        \cite{Le2020}&Wireless cellular network&Mobile users&BS&Training resources&Mobile devices submit their optimal bids for these resources, and
        the winner selection problem is then solved by using primal-dual based greedy algorithm.&SWM, energy cost minimization for sellers, truthfulness, IR and CE\\
        \hline
        \cite{Fan2021}&Edge computing&ENs&Requester&Data resources& The winner and their payment are determined based on the monotone greedy algorithm.& Service valuation maximization for the buyer, utility maximization for sellers, budget
        feasibility, truthfulness, and CE.\\
        \hline
        \cite{Jiao2019}&Wireless FL service market&DOs&FL platform&Data resources& Same as \cite{Fan2021}, but the improvement of social welfare is further studied based on DRL.&
        SWM, IC, IR, and CE\\
        \hline
        \cite{Zhang2021b}&Horizontal FL&Mobile devices&Requester&Data resources&
        The winner selection and payment are determined based on greedy algorithm.&
        Utility maximization for the buyer, IC, IR, budget feasibility, and CE\\
        \hline
        \cite{Lu2021}&MEC&Edge clients&Cloud server& Training resources&
        The optimal bid prices for sellers can be determined using the first-order condition, and the K sellers with lowest bid are selected as the winners.&
        Revenue maximization for sellers\\
        \hline
        \cite{Zhang2019a}&Power demand response&Power users&Power company&Power&The participating users
        are determined based on the price and the expected response rate of users.
        & The total cost minimization for the power company, and fairness\\
        \hline
         \cite{Le2020a}&Cellular wireless network&Users&BS&Computational resources&The optimal uplink transmission power and CPU frequency of sellers can be obtained using iteration algorithm, and then the randomize auction scheme is used to tackle the winner selection problem.& Utility maximization for sellers, social cost minimization for the buyer, truthfulness, IR, and CE.\\
        \hline
        \cite{Zeng2020}&MEC& ENs & Aggregator & Training resources&Given the buyer's scoring rule, sellers determine their optimal bids. Then, the buyer selects K sellers with the highest scores as winners by using Lagrange multiplier method.& Profit maximization for both sellers and buyer, lightweight incentive mechanism, IR, IC, and Pareto efficiency\\
        \hline
        \cite{deng2021fair}&Horizontal FL&Mobile users&FL platform&Training resources&The learning tasks allocation and payment determination are determined by solving the sum of quality of aggregated model updates maximization problem based on the greedy algorithm.& The sum of quality of aggregated model updates maximization, truthfulness, IR and CE\\
        \hline
        \end{tabular}
    \end{table*}

    However, there is a drawback in the approaches mentioned in \cite{Fan2021} and \cite{Jiao2019}, i.e., it needs to obtain the global distribution of all data before calculating EMD, which is impractical under the setting of FL. Therefore, the authors in \cite{Zhang2021b} adopted the reputation mechanism to indirectly represent data quality. Based on this method, they designed an reverse auction-based incentive mechanism (RRAFL), which helps the task requester (i.e., the buyer) choose high-quality and reliable mobile devices (i.e., the sellers) for performing the task. RRAFL is mainly composed of four parts. In the first phase, the comprehensive reputation of mobile devices is calculated. The next phase is winner selection and payment. Based on the reputation value, the candidates calculate their unit reputation bid price, and then they are sorted in a non-decreasing order of their price. The requester select K candidates as winners from the ranking form from front to back which can maximize its utility, and pay to them. The utility of the request is a function of the comprehensive reputation. After the model training, the winners' contribution is measured to reevaluate and update to their reputation value. The experimental results show the effectiveness of RRAFL. In the future, the further study on how to measure the participants’ contributions more reasonably can be considered.

    Paying attention to the same problem as that in \cite{Zhang2021b}, the author in \cite{Lu2021} proposed a cluster-based clients selection method. The system model consists of a cloud aggregation server and multiple edge clients, i.e., SPs. Accordingly, the clients with the similar initial gradient are divided to the same cluster firstly. Second, to overcome the excessive consumption caused by random selection, an auction-based clients selection method is designed in each cluster, in which the clients as bidders provide data services and computing services, and the cloud server as an auctioneer (also the buyer) trains a global model. In the auction, each client determines its optimal bid price that maximize the expected revenue through using the first-order optimal condition. Then, the sever randomly selects a group, and selects K clients with the lowest bid in this group as winners. The minimum local data size among winners as the threshold. The winner selection process is repeated among clients with the data size  greater than the threshold in each group. The simulation results show that the auction-based clients selection scheme can show a faster convergence rate compared to randomly selecting a client from each groups.

    To motivate power users to participate in the interruptible load management, the authors in \cite{Zhang2019a} proposed a multi-attribute sealed auction game to simulate the interaction between the power company and the power users, with the aim of maximizing the power company's revenue and finally decide the list of participating users. The auction model consists of two main phases, i.e., user bidding and user selection. In the first phase, each power user reports the amount of interruption and its quotation to the company. In the second phase, the K-means clustering method is used to cluster the users according to each users' load features.
    Based on the objective of minimizing the total expenditure,
    the company chooses the users from various clusters considering the users' bids and the expected response rate. Meanwhile, the fairness mechanism in the scenario where users have participated in multiple consecutive times is considered.
    The simulation results show that the proposed greedy approach can reduce the expenses of power company and improve the selected probability of small business and residential users. However, the bidding sequence should be considered since it may affect the user selection of long-term demand respond process.

    The system model in \cite{Khan2020} is also considered in \cite{Le2020a}. However, the difference is that the interaction between the BS and users is modeled as a randomized auction where the former is the buyer and the latters are the sellers. In particular, each users determines its bid, consisting of the resources (uplink transmission power and CPU frequency) and training cost, to maximize its utility. The utility of the user is the difference between the reward and the true cost. The optima uplink transmission power and CPU frequency can be obtained using the iteration algorithm. The winner selection problem for the BS is formulated as a social cost minimization problem. To deal with the NP-hard minimization knapsack problem, the framework of randomized auction \cite{Li2017},\cite{Zhong2017} is adopted. The simulation results demonstrate the proposed randomized auction scheme can guarantee the approximation factor of the integrality to the optimal minimum cost. However, the work does not consider the privacy protection of bidders.

    In \cite{Zeng2020}, FMore, a multi-dimensional procurement auction-based incentive mechanism with only one buyer, was proposed with the aim of motivating high-quality ENs with low cost to join in FL. The system model includes an aggregator at the remote cloud, i.e., a buyer, which rents the data, computation and communication resources for FL from ENs i.e., sellers. Specifically, when receiving a bid request with scoring function from the buyer, the ENs decide whether to bid or not according to their available resources. The scoring function is related to the resources quality and sellers' expected payment. If one EN determines to join in FL, it will reply the aggregator the response message (ask) containing the resources quality and the corresponding prices according to sealed-bid auction. The $K$ ENs with the best scores are chosen for participating in collaborative learning, and the payment for these nodes is implemented according to the first-price auction policy. Note that the proposed multi-dimensional auction has multiple winners which is different from \cite{Roy2021}. Extensive simulations show that FMore can speed up federated learning via reducing training rounds by almost 51.3 $\%$, and improve the model accuracy by 28$\%$ for the LSTM model compared with the classic FL scheme, which enables FMore as a lightweight resource auction approach for FL. However, the budget constraint of the aggregator is not considered in FMore, which is left to be extended for the future work. Moreover, for each EN, whether the selected probability should be the same or different remains to be further studied.

    Although the existing works that have applied the reputation mechanism can select the high-quality DOs to join in FL, none of them considers the learning quality of participants, which can significantly affect the performance of FL and the direction of incentive. To bridge this gap, the authors in \cite{deng2021fair} proposed a novel distributed learning system named FAIR, i.e, \underline{F}ederated le\underline{A}rning with qual\underline{I}ty awa\underline{R}eness. Particularly, FAIR integrates the reverse auction-based incentive mechanism design and the quality-aware model aggregation algorithm, which allows it to facilitate precise user incentive and model aggregation. The system mainly consists of three components: the learning quality evaluation, incentive mechanism and model aggregation. Specifically, the historical quality records are firstly leveraged to estimate the current learning quality. Based on the estimated quality, a reverse auction incentive mechanism is then established to encourage mobile users participation, in which the mobile users (the sellers) send their bids to the cloud platform (the auctioneer). To maximize the collective learning quality of all the participants in each iteration, within the limited budget, a learning quality maximization (LQM) problem is formulated. A greedy algorithm based on Myerson's theorem \cite{myerson1981optimal} of truthfulness is designed to solve this NP-hard problem. Thus, the learning task allocation and reward assignment is determined. What's more, a new aggregation algorithm is proposed to filters out non-ideal model updates for the enhancement of the global model. The performance evaluation results show that the performance of the ResNet model can be improved by 68.9$\%$ using FAIR compared with the knapsack greedy and bid price first mechanisms. For the future work, integrating the communication and computation performance into FAIR to enhance its robustness when employed in practical systems can be considered.

\textbf{Summary:} In this section, we introduce the application of auction model for incentive mechanisms in FL. The objective is to maximize the SWM, the revenue of resources providers, and the FL system performance. The reviewed approaches are summarized along with the references in Table \ref{tab:table4} and Table \ref{tab:table5}. As seen, the auction models are mostly used to select high-quality participants to maximize the performance of  FL system. However, the collusion problem is unfavourable for auction-based incentive mechanisms. Thus, it is crucial to develop more effective methods to prevent dishonest bidding. In the next section, we will discuss how to apply the contract and matching theory for incentive mechanism design.


\section{Applications of Contract and Matching Theory for Incentive Mechanism Design in FL}
\label{sec.application.contract.matching}
 Due to the clear decision tree of duties and obligations, contract theory is widely applied in FL systems. For the contract theory based incentive scheme, the MO proposes a list of contracts to DOs. In particular, the MO is not informed about the private type of DOs when designing the contracts. Each DO then proactively selects the appropriate contract type that maximizes its own utility based on the evaluation of their own abilities. Besides, the matching theory is used to match the MO with the DO optimally for the efficient FL training.

    \subsection{Incentive Mechanisms Based on Contract Theory}
    The contract theory-based incentive mechanism aims at maximizing the payoff or utility of the MO. Generally, the problem is to maximize the objective function, i.e., the MO's utility, subject to IR constraint and IC constraint. The first constraint indicates that the payoff gained by the DO under this contract is not less than its reservation payoff if not participating. The second constraint means that the DO's expected payoff is maximized when signing the contract.

    The authors in \cite{Lim2020e} proposed a contract-theoretic task-aware incentive mechanism to model the tradeoff between preferences for service latency and AoI of different training tasks. The problem of training resource trading for the MO having different preferences to the service latency and AoI is considered. The MO acts as the employer to set the contract for the different FL tasks as (reward, the number of update cycles), and the workers as the employees which select one optimal contract to participate in FL. Further, the authors considered to accurately design the reward structure, the update expense, e.g., the energy consumption incurred in data collection and model training is required to be known. This asymmetric information is tackled by using the contract theory. To obtained a feasible contract, the profit maximization problem for the MO is built, subject to IR and IC constraints for the utility of worker. For the future work, the deviation from an ideal cache can be considered and then a caching scheme can be designed to better manage the AoI-service latency tradeoff.

    Considering the challenges when implementing the FL in IoV networks (such as dynamic activities and the limited payment budget), the authors in \cite{Saputra2021} proposed a novel economic framework to address these issues. A multi-principal one-agent contract is designed where the best smart vehicles (SV, as principals) non-collaboratively offer contract agreements (containing information significance and offered payment) to the vehicular service provider (VSP). The VSP (as the agent) is then in charge of optimizing the offered contracts. The payment budget of the VSP is referred as the asymmetric information. Specifically, the VSP perform an SV selection method to decide a set of the best SVs for the FL process according the significance of their current locations and information history. Then, each selected SV can collect on-road information and offer a payment contract to the VSP based on its collected QoI. This contract model is modeled as a profit maximization problem for the VSP and SVs is formulated amid the limited payment budget and common constraints (i.e., IR and IC). To find the optimal contracts for the SVs, the problem is transformed into an equivalent low-complexity problem. The equilibrium solution is the found based on the proposed iterative algorithm. The experiment results demonstrate that the proposed scheme can improve the social welfare of the network up to 27.2 times and convergence 57$\%$ faster compared with those of random and round-robin scheduling methods.

    The economic approach in \cite{Saputra2021} is also applied into the electric vehicle network as proposed in \cite{Saputra2020}. In this network, multiple charging stations (CSs), i.e., principals (employers), firstly implement the energy demand prediction method leveraging FL, and then reserve energy from the smart grid provider (SGP) in advance based on the prediction results to maximize their own profits. The profit of the CS is the difference between the revenue for serving electric vehicles and the energy transfer payments to the SGP. Note that the competition among the CSs and the willingness to transfer energy of the SGP is unknown for each other. The CS's profit maximization problem is modeled as a non-collaborative energy contract problem under the unknown information and the common constraints of the SGP. To find the optimal contracts for the CSs, the method of solution in \cite{Saputra2021} is also used . The simulation results show that the proposed framework can outperform other economic models by 48$\%$ and 36$\%$ in terms of the CSs’ utilities and social welfare (i.e., the total profits of all participating entities) of the network, respectively.

    To simulate the mobile devices with high-quality data to participate in FL, the authors in \cite{Kang2019a} designed a contract theory-based incentive mechanism, where mobile devices (DOs) are the employees and the monopolist operator (i.e., task publisher) is the employer. The task publisher design specific contracts for different types of DOs with different levels of data quality to maximize its profits, which is a function of the time of the global iteration and the reward paid to DOs. Under the constraints of IR and IC, the optimal CPU resources and the corresponding reward can be derived using the substitution method and the convex optimization tools. The simulation results show that the the task publisher can obtain higher profit using the proposed contract-based method compared to the Stackelberg game method in \cite{Kang2019b}. Based on this approach, the reputation mechanism based on blockchain is introduced in \cite{Kang2019} to measure the reliability and trustworthiness of the mobile devices, so as to motivate high-reputation DOs with high-quality data to join in model learning. To further improve the accuracy of reputation computation, more weight parameters can be taken into consideration.

    Unlike the contract-based incentive mechanisms that consider single-dimensional privacy information, the work in \cite{Ding2020} proposed the multi-dimensional contract theoretic approach which considers the user's two-dimensional privacy information including training cost and communication delay. The interaction between the FL serve and users is modeled as a labor market, in which the former act as the employer and the latter act as the employees.
    The key contribution of this approaches is summarizing users’ multi-dimensional private information into a one-dimensional criterion (i.e., the server's preference on different user' types) that allows a complete order of users. Three scenarios are considered including complete informational scenario, weakly incomplete information scenario, and strongly incomplete information scenario is considered. In the first scenario, the serve knows each user's type, the contract design is to minimize the server's cost with the IR constraint. In the second scenario, the same objective with IR and IC constraints is considered since the server does not know which user belongs to which type. In the third scenario, the contract design is to minimize the expect cost of the server, where the objective function is similar to that in the second scenario excepted the expectation part. The simulation results show that all the designed contract in the three information scenarios can achieve up to 73.72$\%$ cost reduction of uniform contract when the number of users is large.

    The approach in \cite{Ding2020} assumed that the data of user is IID, in fact, data distribution may have an impact on the server's cost. Therefore, the authors further studied the case where users have non-IID data in \cite{Ding2021}. And the EMD metric is used again. The author derived the server's optimal contract with the non-IID data under complete and weakly incomplete information scenarios, respectively. Compared with the case IID in which the server only selects the most preferred type under complete information, in the non-IID case, the server may choose more than one type. Meanwhile, they find that weakly incomplete information does not increase the server’s cost (comparing with the complete information scenario) when training data is IID, but it in general does when data is non-IID. The experimental results of the non-IID case show that the server's cost in the proposed  mechanism is lower than that in optimal uniform contract (OUC), reverse multi-dimensional auction mechanism (RMA) \cite{Jiao2019} and Stackelberg game mechanism (SBG) \cite{Feng2019}. For the future work, deriving the optimal contract for the strongly incomplete information scenario, and expanding existing researches to a general scenario with multiple competitive servers can be considered.

    The authors in \cite{Lim2020d} leveraged on multi-dimensional contract theory to design a trustworthy incentive mechanism for the UAV-enable IoV network. The system model consists of a MO, i.e., the employer, perform a time-sensitive task, and multiple UAVs, i.e., the employees, implement sensing task and transmit the collected data to the MO. The objective of this incentive is to maximize the profit for both the MO and UAVs. The profit of UAV is the difference between the contractual rewards and energy cost, and the profit of the MO is the total revenue obtained from UAV and the reward expenses. In this incentive scheme, a UAV is categorized into a multi-dimensional type-$(x, y, z, q)$, where $x$ is traversal cost type, $y$ is sensing cost type, $z$ is computation cost type, and $q$ is transmission cost type. The proposed contract design includes two phase, i.e, multi-dimensional contract design and traversal cost compensation. In the first phase, the sensing cost and computation cost is transformed into a single-dimensional contract design problem, same as the method in \cite{Ding2020}. In the second,  a fixed compensation is added to derive the final contract bundles by considering additional traversal cost and transmission cost. After obtaining the optimal contract, the matching-based UAV-subregion assignment using the Gale-Shapley algorithm \cite{Lim2021} is considered to match the lowest cost UAV to each subregion, i.e., target sensing region. The simulation results demonstrate the IC property of the proposed contract design and show the efficiency of the matching.

    \subsection{Incentive Mechanisms Based on Matching Theory}
    Matching theory is widely used in the real market for its optimal two-side matching rule. This property is beneficial for the worker selection of the requester of the FL service.

    The author in \cite{Fantacci2020} considered the network slicing placement strategy to offering customize service for different users' heterogeneous requirements. In particular, the network slice controller (NSC), i.e., the seller, collects the requests from end users (EU), i.e., buyers, and then deploys the virtual network functions (VNFs) to provider different communication services for each of them. The proposed VNFs placement policy includes three modules, i.e., FL module, EUs behavior module, and VNFs placement module. The first module is deployed among NSC and EUs designed to predict EUs' VNFs demand information for each type of service. Based on the prediction results, the network slice controller provide and place the corresponding VNF to the area of EUs. A two-phase placement strategy based on the matching theory is proposed to maximize the SP's revenue. The SP revenue is the difference between the price payed by the EU based on its level of quality of experience and the cost borne by the SP in provisioning and placing the VNF in the service area. As shown in the simulation results, the proposed matching theory-based assignment approach can reach a superior revenue compared with the method based on Kolkata game \cite{Chakrabarti2009} and potential game \cite{Yao2016}.

\textbf{Summary:} In this section, we discuss the applications of contract and matching theory for incentive mechanism design in FL. The objective is to maximize the seller's utility or revenue, the buyer's profit, and to minimize the latency of FL training. The related works are summarized along with the references in Table \ref{tab:table6}. As seen, the contract design mostly depend on the traditional optimization method, in the future, the DRL-based contract design can be considered to better adapt to the changing user types.

\begin{table*}[!t]
        \newcommand{\tabincell}[2]{\begin{tabular}{@{}#1@{}}#2\end{tabular}}
        \centering
        \scriptsize
        \caption{Applications of Contract and Matching Theory for Incentive Mechanism Design in FL}
        \label{tab:table6}
        \begin{tabular}{|m{0.4cm}<{\centering}|m{1.6cm}<{\centering}|m{1.1cm}
        <{\centering}|m{1.1cm}<{\centering}|m{1.2cm}<{\centering}|m{5.7cm}
        <{\centering}|m{3.7cm}<{\centering}|}
        \hline
        \multirow{2}{*}{\bf Ref.} & \multirow{2}{*}{\bf Scenario} & \multicolumn{3}{c|}{\bf Market Structure} & \multirow{2}{*}{\bf Mechanism}& \multirow{2}{*}{\bf Objective}\\
        \cline{3-5}
        && {\bf Seller} & {\bf Buyer} & {\bf Item} &&\\
        \hline\hline
        \cite{Lim2020e}&Smart industries&Workers&MO&Training resource&For the tradeoff between the service latency and AoI, the frequency of data update are stipulated in the optimal contract&Profit maximization for the MO, IR and IC \\
        \hline
        \cite{Saputra2021}&IoV network&Vehicles&Vehicular SP&Vehicles' local training results&The optimal contract problem is converted a equivalent low-complexity problem, and then the equilibrium solution for vehicles is obtained using the iteration algorithm.&Profit maximization for both buyer and sellers, QoI improvement for sellers, IR and IC\\
        \hline
        \cite{Saputra2020}&Electric vehicle networks&SGP&CSs&Power&Same as \cite{Saputra2021}&Profit maximization for buyers, IC,and IR\\
        \hline
        \cite{Kang2019a} \cite{Kang2019}&Mobile network&Mobile devices&Task publisher&Data resources&
        The optimal contributed data size and reward are obtained using the substitution method.&Profit maximization for the buyer, guarantee latency requirement, IR and IC\\
        \hline
        \cite{Ding2020} \cite{Ding2021}&Efficient FL&Users&The central server&Data resources
        &The optimal contract can be derived by converting the two-dimensional information into a single-dimensional criterion.&Payoff maximization for sellers and cost minimization for the buyer, guarantee latency requirement, IR and IC\\
        \hline
        \cite{Lim2020d}&UAV-enable IoV network&UAVs&MO&Sensing task&The optimal contract can be obtained using the same method in \cite{Ding2020}, and the lowest cost UAV is allocated each subregion based on the Gale-Shapley algorithm. & Profit maximization for both sellers and the buyer, IR and IC.\\
        \hline
        \cite{Fantacci2020}&Network slicing&NSC&EUs&Network slice service&The NVFs placement strategy is modeled as a two-phase matching game&Revenue maximization for sellers, and guarantee the QoE requirement of the buyer\\
        \hline
        \end{tabular}
    \end{table*}

\section{Summary, Challenges, and Future Research Directions}   
\label{sec.summary}
Different approaches reviewed in this survey show the effectiveness of economic and game theoretic models for the incentive mechanism design for the FL systems. Evidently, economic and game theoretic models have been widely applied to design various incentive scheme in different scenario. However, incentive mechanism design is in its infancy. Thus, there are still open issues and new research directions on incentive mechanisms for FL as discussed in the following.

\emph{1) Online-Auction based Mechanism Design:} Existing monetary incentive schemes based on traditional auction method mostly are in offline, i.e., the auction starts only if a sufficient number of bidders must be available. For example, in \cite{Zeng2020}, the model aggragator starts to determine the winner when the number of the bids from the ENs is sufficient or a predefined users’ bids have to be synchronized. In such an auction, each bidder may need to wait for a long time even if the bidder does not win the auction. This discourages the bidder to join in the FL platform. Different from the offline auctions, online auctions allow the auctioneer or the seller to makes decisions, i.e., on winner and prices, upon a bidder joins the auction. The online auction can break the time and space constraints and save costs. Some recent works, e.g., \cite{Zhang2014} and \cite{Zhao2014}, have investigated the online auction for mobile crowdsensing. Thus, the online auctions is a promising solution for the incentive mechanism design for FL.

\emph{2) Advanced Contribution Evaluation Methods:} To evaluate the contribution of the data owners, Shapley value have been commonly used. Based on the concept of the SV, many works have been done such as the federated SV \cite{Wang2020} based on local model updates and contribution index \cite{Song2019} based on the intermediate results of the training process of FL. The existing works demonstrate that the SV method in FL setting has comparable performance. In addtion, a computation-and communication-efficient estimation method is proposed in \cite{nishio2020estimation}. The main focus of these works is to reduce the time complexity of the computing SV. Note that they assume that a trusted server will honestly evaluate the contribution of each participant. To establish a transparent evaluation mechanism, blockchain-based SV \cite{Ma2021} is adopted to overcome the monopolistic behavior. However, all of the aforementioned works do not consider the adversarial behaviours of the participants, which may affect the the Shapley value calculation and lead to the unfairness. More research works need to be investigated for designing comprehensive and transparent contribution evaluation mechanisms.

\emph{3) Incentive Mechanism with Privacy Protection:} Existing incentive mechanisms largely ignore the protection of privacy information, e.g., individual preferences, of data owners. What’s worse, with the most current incentive mechanisms, data owners who participate in the bidding process directly reveal their privacy information. Meanwhile, the data owners who lose in the bidding process receive no compensation at all for their privacy revelation, which may reduce the participation enthusiasm of data owners. To protect data owners’ privacy while maintaining high utility of trained models, some works that consider the privacy problem have been done. The works in \cite{Pejo2019}, \cite{zheng2021incentive} proposed the game theoretic models for the FL market, in which the data owners can obtain compensation according to their privacy loss quantified by local differential privacy. However, these frameworks may not be suitable for some specific learning tasks. Thus, more research works need to be investigated for both incentiving data owners and protecting their privacy.

\emph{4) Comprehensive Incentive Scheme:} The most of current incentive mechanisms aim to optimize a single objective. For example, the works \cite{Zhan2020a, Khan2020, Pandey2020} aim to select the optimal participants for maximizing the performance of the FL system. Also, some works related to the Shapley value aim to evaluate the contribution of the participants to guarantee the fairness. In fact, multiple objectives should be considered in the FL. For example, the participant selection and the contribution evaluation should be jointly investigated to maximize both the FL performance while guaranteeing the fairness. Therefore, designing incentive mechanisms with multiple objectives are needed.

\emph{5) Integration of Incentive Mechanism and Cutting-edge Technologies:} Recently, DRL-based incentive mechanism designs have attracted great attention and achieved good results, such as \cite{Zhan2020}, \cite{Zhan2020a}. Thus, some cutting-edge technologies, such as graph neural networks, generative adversarial networks, multi-agent reinforcement learning, etc., might find its potential application in the incentive design of new scenarios like mobile edge computing and 5G/B5G.

\emph{6) Multiple Incentive Schemes Co-Existence:} Different participants might prefer different kinds of incentives, however the current FL systems only adopt one single incentive mechanism (i.e., monetary incentive mechanism). Multiple incentive schemes allow different participants simultaneously obtain different types of rewards. This idea faces some challenges, such as how to determine what type of reward should be offered to which participant, and who make this decisions. In addition, when a new participant arrives, the FL platform should be able to flexibly design a new scheme that is individualized to the new participant according to its preferences.

\section{Conclusion}    \label{sec.conclusion}
This paper has presented a comprehensive survey on the applications of economic and game theories to incentive mechanism design in FL. Firstly, we have reviewed the fundamental and background of FL and incentive mechanism design. Then, we have introduced fundamentals of various economic and game models with the aim to understand the motivations of using the economic and game models in incentive mechanism design of FL. After that, we have provided detailed reviews, analyses, and comparisons of the economic and game theoretic approaches in designing variety of incentive mechanisms of FL. Finally, we have discussed some open issues and future research directions.

\bibliographystyle{IEEEtran}
\bibliography{cite}

\end{document}